\documentclass[extra,mreferee]{gji}
\usepackage{timet,amsmath,graphicx,amssymb}

\title[Seismoelectric effects of 1-D heterogeneities]
      {An analytical study of seismoelectric signals produced by 1D
        mesoscopic heterogeneities} 

\author[Monachesi et al.]  {Leonardo B.  Monachesi$^1$, J.  Germ\'an Rubino$^2$,
  Marina   Rosas-Carbajal$^2$, \and  Damien   Jougnot$^{2,3}$,  Niklas Linde$^2$,
  Beatriz   Quintal$^2$ and Klaus  Holliger$^2$  \\ $^{1}$ CONICET -  Facultad de
  Ciencias  Astron\'omicas y  Geof\'isicas, Universidad  Nacional  de La
  Plata, \\ La Plata,  Argentina\\  $^{2}$  Applied  and  Environmental  Geophysics Group, Institute of Earth Sciences, University of Lausanne,\\ CH-1015 Lausanne, Switzerland   \\  $^{3}$  Now at CNRS, UMR 7619, METIS, F-75005 Paris, France}

\begin{document}
\label{firstpage}
\maketitle
%%%%%%%%%%%%%%%%%%%%%%%%%%%%%%%%%%%%%%%%%%%%%%%%%%%%%%%%%%%%%%%%%%%%%%
%%%%%%%%%%%%%%%%%%%%%%%%%%%%%%%%%%%%%%%%%%%%%%%%%%%%%%%%%%%%%%%%%%%%%%

\noindent
This paper has been accepted for publication in Geophysical Journal International:

\noindent
L. B. Monachesi, J. G. Rubino, M. Rosas-Carbajal, D. Jougnot, N. Linde, B. Quintal, K. Holliger (2015) An analytical study of seismoelectric signals produced by 1D mesoscopic heterogeneities, \textit{Geophys. J. Int.}, 201, 329 - 342, doi: 10.1093/gji/ggu482.

\begin{summary}
The presence of mesoscopic heterogeneities in fluid-saturated porous rocks can produce measurable seismoelectric signals
due to wave-induced fluid flow between regions of differing 
compressibility. The dependence of these signals on the petrophysical
and structural characteristics of  
the probed rock mass remains largely unexplored.
In this work, we derive an analytical solution to describe the seismoelectric response
of a  rock sample,  containing a horizontal layer at its center, that is
subjected  to an oscillatory    compressibility test.   
We then adapt this general solution to compute the  seismoelectric signature of a 
particular case related to a sample that is permeated by a horizontal fracture located
at its center. Analyses of the general and particular solutions are performed
to study the impact of different  petrophysical and structural parameters 
on  the  seismoelectric response. We find that the amplitude of the seismoelectric signal is directly 
proportional to the applied stress,  to the Skempton coefficient contrast between the host rock 
and the layer, and to a weighted average of the effective excess charge of 
the two materials.  
Our results also demonstrate that 
the frequency at which the maximum electrical potential amplitude prevails
does not depend on the applied stress or the Skempton coefficient contrast.
In presence of strong permeability variations, this frequency is rather
controlled by the permeability and thickness of the less permeable material. 
The results of this study thus indicate that seismoelectric measurements can
potentially be used to estimate key mechanical and hydraulic rock properties
of mesoscopic heterogeneities,
such as  compressibility, permeability, and fracture compliance. 

\end{summary}

\begin{keywords}
Hydrogeophysics  --  Electrical properties  --  Fracture  and flow  --
Permeability and porosity.

\end{keywords}

%%%%%%%%%%%%%%%%%%%%%%%%%%%%%%%%%%%%%%%%%%%%%%%%%%%%%%%%%%%%%%%%%%%%%%
\section{Introduction}
%%%%%%%%%%%%%%%%%%%%%%%%%%%%%%%%%%%%%%%%%%%%%%%%%%%%%%%%%%%%%%%%%%%%%%

Free electrical charges at the mineral surfaces of  fluid-saturated porous
rocks are responsible for an electrical  double layer  (EDL)
in the  pore  space surrounding  the  solid grains.   The EDL  is
characterized  by an  electrical excess  charge that counter-balances the
surface  charges \citep{Hunter1981}. Pore fluid flow in response to a propagating seismic wave
\citep{biot_high,biot_low} exerts a drag on this excess charge, thus resulting in an
electrical source current generally referred to as streaming current.
This in turn results in  an  electrical potential  distribution,
which  can  be measured  in  laboratory  and  field experiments. The
thus generated electric field depends on a range of important petrophysical parameters,
such as the porosity, the permeability, and the type  of saturating  fluid. This
has fostered an increased interest in this physical process, 
commonly referred to as seismoelectric  conversion
\citep[e.g.,][]{thompson1993geophysical,jouniaux2011electrokinetic}.

\cite{pride94} developed the  theoretical basis of the seismoelectric
phenomenon  by  coupling  Biot's  (\citeyear{biot_high,biot_low})  and
Maxwell's equations  through a volume-averaging  approach. This theory
predicts  two  kinds of  seismoelectric conversions:  (1) the
coseismic field and (2) the interface response. The coseismic field is
a consequence  of the wavelength-scale relative fluid flow  associated with the  
passing seismic wavefield,  which in turn generates a  streaming current
and thus an  electric field. This field travels
with the seismic wave, even in the case of homogeneous media,  but is
largely limited to the wave's support volume
\citep[]{pride-haartsen1996}. Conversely, the  interface  response
occurs when a seismic perturbation strikes on an interface in terms
of mechanical or electrical properties. In this case,
a variation in the streaming current distribution arises, breaking
the  symmetry  in  the   charge  separation. This generates
an electromagnetic (EM) signal that propagates independently of the seismic
wave and can be measured outside the support volume of the wave,
thus allowing for the remote detection of geological interfaces.
These signals are  highly sensitive to the fluid pressure
gradients in the vicinity of the interface. Thus, proper modeling of
wave conversions at  interfaces, and in particular of Biot slow waves,
is  critical  for the  accurate  modeling  of seismolectric  interface
responses \citep{pride-garambois2002}.

The propagation of seismic waves through a
medium containing mesoscopic heterogeneities, that is, heterogeneities
having sizes larger than the typical pore scale but smaller than the
prevailing wavelengths, can produce significant oscillatory fluid
flow, generally referred to as wave-induced fluid flow, between the different 
regions composing the heterogeneous medium
\citep[e.g.,][]{muller-et-al10}.
Due to the differing  elastic compliances of the various regions, the stresses associated with
the seismic perturbation  produce a  pore fluid  pressure gradient
and, consequently, fluid flow. The energy dissipation related
to this  phenomenon is  considered to  be one of  the most  common and
important  seismic attenuation mechanisms in the shallower
parts  of the crust  \citep[e.g.,][]{muller-et-al10}.
Given that the
amount of fluid flow produced by this phenomenon can be significant, a
potentially important interface-type seismoelectric signal is also expected to arise.
However, the  nature and importance of  the seismoelectric effects related
to mesoscopic heterogeneities remain largely
unexplored. An interesting study of this kind was presented by
\cite{haartsen-pride1997} who modeled  the seismoelectric response
of  a single sand layer having a thickness much smaller than the predominant 
seismic wavelength and embedded in a less permeable medium.
They observed that while the seismic amplitudes recorded at the surface were 
very small due to destructive interferences, the converted EM
amplitudes  were significantly enhanced compared  to the case of a 
thick sand layer. More  recently, \citet{jougnot2013seismoelectric}
proposed a numerical framework to study the seismoelectric response of
a rock sample containing mesoscopic fractures subjected to an oscillatory
compressibility test \citep{rubino-et-al09,rubino_et_al_jasa13} and found seismoelectric signals that would be measurable for typical laboratory setups
\citep[e.g.,][]{tisato2013measurements}.
These findings are important not only as the seismoelectric responses
of most geological environments are expected to contain  a
component related to fluid flow at mesoscopic scale, but also because
they open the perspective of developing seismoelectric spectroscopy as
a  novel  laboratory method  for  characterizing heterogeneous  porous
media. \cite{jougnot2013seismoelectric} suggest that a better understanding of the 
role played by mesoscopic heterogeneities could help to improve 
some of the practical aspects of the seismoelectric method, such as the notoriously 
high noise levels generally observed in the measurements.

In this work, we present an analytical approach to study the origin of the 
seismoelectric signal produced by mesoscopic heterogeneities.
We first derive a general analytical solution for the seismoelectric response
of a homogeneous rectangular rock sample containing a horizontal layer at
its center and subjected to an oscillatory compressibility test.
Following \citet{brajanovski_et_al05}, we then adapt the analytical solution 
to the particular case of a sample containing a central layer having thickness and 
dry-frame elastic moduli tending to zero in conjunction with  porosity tending
to one. This particular
solution represents the response of a sample permeated by a horizontal 
fracture at its  center. Finally, we employ these two solutions
to explore the roles played by different petrophysical and structural properties
in the seismoelectric signatures of heterogeneous rocks. This analysis 
may, in turn, help to identify which parameters  could be retrieved from seismoelectric 
measurements.

%%%%%%%%%%%%%%%%%%%%%%%%%%%%%%%%%%%%%%%%%%%%%%%%%%%%%%%%%%%%%%%%%%%%%%
\section{Methodology}
%%%%%%%%%%%%%%%%%%%%%%%%%%%%%%%%%%%%%%%%%%%%%%%%%%%%%%%%%%%%%%%%%%%%%%

%%%%%%%%%%%%%%%%%%%%%%%%%%%%%%%%%%%%%%%%%%%%%%%%%%%%%%%%%%%%%%%%%%%%%%
\subsection{Oscillatory compressibility test}

\begin{figure}
\begin{center}
\includegraphics[angle=0,width=0.4\textwidth]{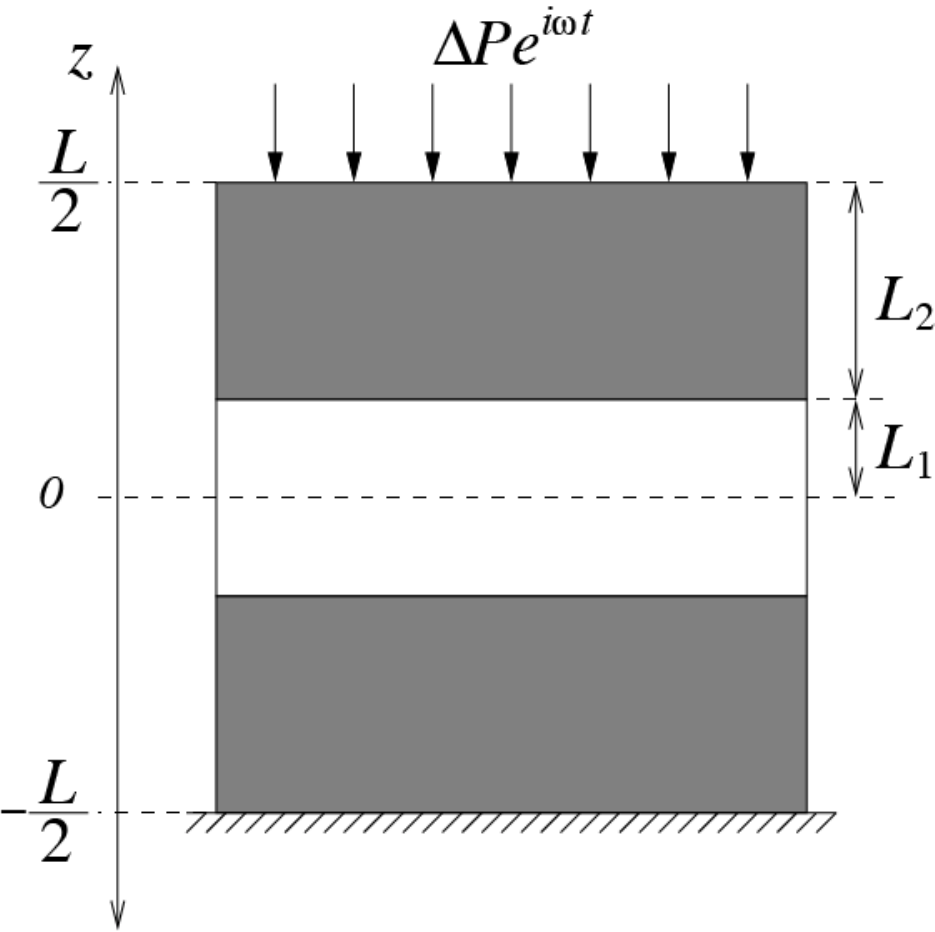}
\caption{Schematic representation of  a rectangular porous rock sample
  containing a horizontal  layer (white rectangle) embedded
  between  two  identical regions (gray  upper  and  lower
  rectangles).   The   sample   is   subjected   to   an   oscillatory
  compressibility test.}
\label{three_layers}
\end{center}
\end{figure}

We consider a rectangular fluid-saturated porous rock sample
containing a horizontal layer located at its center (Fig.~\ref{three_layers}). 
The two regions  embedding this mesoscopic  heterogeneity  are
identical.  For simplicity,  we choose the center  of the sample
as the  origin of  the $z$-axis and,  therefore, the positions  of the
upper  and  lower boundaries  of  the  heterogeneity  are $z=L_1$  and
$z=-L_1$, respectively.  The thicknesses of the two embedding regions 
constituting the host rock are
$L_2$ and the total thickness of the sample is $L=2(L_1+L_2)$.

The sample is subjected  to a time-harmonic compression of the form $\Delta P
e^{i\omega t}$ at its top boundary, with $\omega$  being  the  angular
frequency, $t$ the time, and $i=\sqrt{-1}$
\citep[][]{rubino-et-al09,rubino_et_al_jasa13}.  We  impose  that  the
solid phase is not allowed to move on the bottom boundary, nor to have horizontal
displacements at the lateral boundaries and the pore fluid cannot flow into or
out of the sample. For the considered geometry and boundary conditions,
the  problem  to  solve is 1D.
We consider frequencies $\omega$ smaller than Biot's critical
frequency $\omega_B$ \citep[e.g.,][]{biot62}
\begin{equation}
\omega_B=2\pi f_B=\frac{\phi\eta}{\kappa\rho_f},
\end{equation}
where  $\phi$ is the porosity, $\kappa$ the permeability, 
$\rho_f$ the density  of the  pore fluid, and $\eta$ the fluid viscosity.  In this frequency range,
viscous boundary layer effects are negligible and thus  
we can solve  Biot's (\citeyear{biot_consol})
consolidation equations to obtain the response of the sample. In the
1D space-frequency domain these equations can
be written as
\begin{equation}
\frac{\partial \tau}{\partial z}=0,\label{ecs_motion_a}
\end{equation}
\begin{equation}
i\omega \frac{\eta}{\kappa}w=-
\frac{\partial p_f}{\partial z},\label{ecs_motion_b}
\end{equation}
where $\tau$ and $p_f$ denote the total stress and the fluid pressure,
respectively, whereas $w$ is the relative fluid-solid displacement.
Equation~(\ref{ecs_motion_a}) represents the stress equilibrium within the sample,
whereas Eq.~(\ref{ecs_motion_b}) is Darcy's law.   
These two equations are coupled through the 1D stress-strain relations
\begin{equation}\label{mod.2a}
\tau=H\frac{\partial u}{\partial z}+\alpha  M
\frac{\partial w}{\partial z},
\end{equation}
\begin{equation}\label{mod.2b}p_f=  - \alpha  M \frac{\partial u}{\partial z}- M
\frac{\partial w}{\partial z}.
\end{equation}
In these equations, $u$ denotes  the average displacement of the solid
phase and the coefficients $H$, $M$ and $\alpha$ are given by
\begin{equation}
H=\lambda+2\mu,\label{H_c}
\end{equation}
\begin{equation}
M=\left[\frac{\alpha-\phi}{K_s}+\frac{\phi}{K_f}\right]^{-1},
\label{K_av}
\end{equation}
\begin{equation}
\alpha=1-\frac{K_m}{K_s},\label{alpha}
\end{equation}
where $K_m$, $K_s$, and $K_f$ are the bulk
moduli   of  the  solid   matrix,  solid   grains,  and   fluid  phase,
respectively.   Moreover,  $\mu$ is  the  shear  modulus  of the  bulk
material, which is  equal to that  of the dry matrix,
and $\lambda$ is the Lam\'e constant, given by
\begin{equation}
\lambda=K_m+M\alpha^2-\frac{2}{3}\mu.
\end{equation}

%%%%%%%%%%%%%%%%%%%%%%%%%%%%%%%%%%%%%%%%%%%%%%%%%%%%%%%%%%%%%%%%%%%%%%
%%%%%%%%%%%%%%%%%%%%%%%%%%%%%%%%%%%%%%%%%%%%%%%%%%%%%%%%%%%%%%%%%%%%%%

For a homogeneous medium, combining Eqs. (\ref{ecs_motion_a}) and (\ref{mod.2a}), as well as 
Eqs. (\ref{ecs_motion_b}) and  (\ref{mod.2b}) leads to
\begin{equation}\label{aux1}
\frac{\partial^2 u}{\partial z^2}=-\frac{\alpha M}{H}\frac{\partial^2 w}{\partial z^2},
\end{equation}
and
\begin{equation}\label{aux2}
i\omega\frac{\eta}{\kappa}w=\alpha M\frac{\partial^2 u}{\partial z^2}+M\frac{\partial^2w}{\partial z^2},
\end{equation}
respectively. Next, substituting Eq.  (\ref{aux1}) into Eq. (\ref{aux2}), results in
\begin{equation}\label{diffusion}
i\omega w=D\frac{\partial^2w}{\partial z^2}.
\end{equation}
Equation~(\ref{diffusion}) is a diffusion equation, with the diffusivity $D$ given by
\begin{equation}\label{diffusivity}
D=\frac{\kappa N}{\eta},
\end{equation}
where 
$N=M-\alpha^2M^2/H$.

The spatial scale at  which wave-induced fluid flow is significant is determined
by the diffusion length
\begin{equation}\label{diff_length}
L_d \equiv \sqrt{D /\omega}.
\end{equation}
From Eqs. (\ref{diffusivity}) and (\ref{diff_length}) it is clear that
this length increases with increasing permeability and decreasing
viscosity. When the  diffusion length is of similar size as the
characteristic  size  of the heterogeneities, 
$a_{meso}$, a characteristic frequency can be defined as
\begin{equation}\label{w_c}
f_c\approx \frac{D}{2 \pi \ a_{meso}^2}, 
%\omega_{c} \approx D / a_{meso}^2,
\end{equation}
which represents  the limit between two types  of mechanical behaviors
in response  to the oscillatory  compression.  For frequencies  $f \ll
f_c$, the diffusion  lengths are much larger than  the typical size of
the  heterogeneities.   Correspondingly,  there  will be  enough  time
during  each oscillatory  half-cycle for  the pore  fluid  pressure to
equilibrate  at  a  common  value.  Thus,  this  low-frequency  regime
represents a relaxed state.  For frequencies $f \gg
f_c$, the diffusion lengths are very  small compared to the size of the
heterogeneities and, hence, there is no  time for interaction between the pore
fluids of  the different  parts of  the rock.  In  this case,  the pore
pressure  is  approximately constant  within  each heterogeneity  and,
consequently,  the  high-frequency   regime  is  associated  with  an
unrelaxed  state.  For intermediate  frequencies, $f \approx f_c$,
significant fluid flow can be induced   by   an oscillatory stress field
\citep[e.g.,][]{muller-et-al10}. From Eqs.~ (\ref{diffusivity}) and (\ref{w_c})
we notice that the frequency  range  where significant fluid flow can occur shifts 
towards lower frequencies for decreasing permeability, increasing fluid viscosity, 
or increasing size of the heterogeneities.

The general solution of Eq.~(\ref{diffusion}) for a homogeneous medium is given by
\begin{equation}
w(z)=A e^{-k z}+B e^{k z},\label{w_gen_sol}
\end{equation}
where 
\begin{equation}\label{k}
k=\frac{\sqrt{i}}{L_d},
\end{equation}
and   $A$  and   $B$  being   complex-valued  constants.    From  Eq.
(\ref{mod.2a}),  given  that  $\tau$  is  spatially constant in  virtue  of  Eq.
(\ref{ecs_motion_a}), the solid displacement can be expressed as
\begin{equation}
u(z)=-\beta w(z) +\frac{\tau}{H}z+C, \label{u_s_gen_sol}
\end{equation}
where $\beta\equiv\alpha M/H$ is the 1D Skempton coefficient (Appendix A)
and $C$  is an additional complex-valued  constant.  Eqs.
(\ref{w_gen_sol})  to  (\ref{u_s_gen_sol})  then constitute  the  general
solutions   of  Eqs.   (\ref{ecs_motion_a})   to  (\ref{mod.2b})   for
homogeneous media.

Since the generation  of the seismoelectric signal is  produced by the
relative fluid velocity  field, $i\omega w$, we
seek an analytical  expression for this
parameter. The  solid displacement  $u$ always appears  as a derivate with
respect    to    $z$   in    Eqs.    (\ref{ecs_motion_a})    to
(\ref{mod.2b}). This means that if $\left(w(z),u(z)\right)$ constitute
a  solution for such  equations, then  $\left(w(z),u(z)+\chi \right)$,
with $\chi$ being  a constant, is also a  solution. Hence, 
the  solution  for   $w$  is  independent  of  the  solid
displacement   value   imposed  at   the   bottom   boundary  of   the
sample. Correspondingly, without changing the solution for $w$, we can
 modify the  boundary condition  imposed on  $u$ at  the bottom
boundary to  produce $u=0$ at  $z=0$. Under this condition,  and since
$\tau$ is  constant within the sample  (Eq.  (\ref{ecs_motion_a})), it
is clear  that the  geometry and the  imposed boundary  conditions are
symmetrical with respect to  $z=0$ and, therefore, the resulting fluid
flow profile should also be  symmetrical with respect to the center of
the sample.  Hence,  we can simply solve Eq.   (\ref{diffusion}) in the
upper half  of the sample  shown in Fig.~\ref{three_layers}  under the
following boundary conditions
\begin{eqnarray}
\tau&=& -\Delta P,\quad \ \ \ z=L/2,\label{BC_1}\\
w&=&0,\quad \ \ \ \ \ \ \ \ \ z=L/2,\label{BC_2}\\
w&=&0, \quad \ \ \ \ \ \ \ \ \  z=0,\label{BC_3}\\
u&=&0, \quad \ \ \ \ \ \ \ \ \  z=0.\label{BC_4}
\end{eqnarray}
According  to  Eqs.  (\ref{w_gen_sol})  and  (\ref{u_s_gen_sol}),  the
general solutions in  the upper half  of the
sample can be expressed as
\begin{eqnarray}
w_j(z)&=&A_j   e^{-k_jz}+B_j e^{k_jz},\\
u_j(z)&=&-\beta_jw_j(z)-\frac{\Delta P}{H_j}z+C_j,\qquad\qquad j=1,2,
\end{eqnarray}
where $\tau$ has been replaced  by the imposed stress $-\Delta P$.
The subscripts  $1$ and $2$  denote the parameters  of 
the lower  ($0\le z \le  L_1$) and upper  ($L_1\le z \le  L/2$) layers
composing the upper half of the sample, respectively. The six unknowns
$A_j$,  $B_j$,  and $C_j$ ($j=1,2$)  are  determined  by  imposing the
conditions given by Eqs. (\ref{BC_2}) to (\ref{BC_4})  and the continuity  of $u$, $w$,
and  $p_f$ at  the interface  $z=L_1$. Taking these six conditions into account
and the fact that the relative fluid  displacement $w$ is
an odd function, it can be shown that
\begin{equation}\label{w_solution}
w(z)=
\begin{cases}
\begin{array}{lcl}
 sgn(z)   A_{1}\left(   e^{-k_1\lvert  z\rvert}-e^{k_1\lvert   z
  \rvert} \right), \quad  \quad \quad \quad \quad \quad  0 \leq \lvert
z\rvert   \leq   L_{1},\\   sgn(z)  A_{2}\left(   e^{-k_2\lvert
  z\rvert}-e^{-k_2(L-\lvert z\rvert)} \right), \quad \quad \quad L_{1}
\leq \lvert z\rvert \leq L/2,
\end{array}
\end{cases}
\end{equation}
where  $sgn$  is the  sign  function.   The parameters $A_1$  and
$A_2$ are  given by 
\begin{equation}\label{A_1}
A_1=\left(e^{-k_{1}L_{1}}-e^{k_{1}L_{1}}\right)^{-1}
\frac{\Delta P   
\left(\beta_1-\beta_2\right)}{\sum_{j=1}^{2} N_j k_j \ coth (k_jL_j)},
\end{equation}
\begin{equation}\label{A_2}
A_2=\left(e^{-k_{2}L_{1}}-e^{-k_{2}(L-L_{1})}\right)^{-1}
\frac{\Delta P \left(  
\beta_1-\beta_2
\right)}{\sum_{j=1}^{2} N_j k_j \ coth (k_jL_j)}.
\end{equation}

In the  following subsection, the above expressions for the relative
fluid displacement are used to  infer the distribution of the electrical potential
within the  sample in  response  to the  applied
oscillatory compression.

%%%%%%%%%%%%%%%%%%%%%%%%%%%%%%%%%%%%%%%%%%%%%%%%%%%%%%%%%%%%%%%%%%%%%%
\subsection{General solution: seismoelectric response of a rock sample containing a horizontal layer}

%The source of the seismoelectric conversion is related to the presence
%of  an EDL  developing in  the pore  water
%around the  solid grains.  This  EDL consists of an  electrical excess
%charge $\bar{Q}_{v}$  in the  pore water solution  (i.e., counterions),
%which  counterbalances the  charges at  the mineral  surface.   
When a relative  displacement  between  the  pore fluid  and  the solid frame 
occurs in response to the applied oscillatory compression,
a  drag on the electrical  excess charges of  the EDL
takes place. This, in turn, generates a
source or streaming current density $j_s$.
Since the distributions of both the excess charge and
 the microscopic relative velocity of the pore fluid are highly dependent on their distance 
 to  the  mineral grains, not  all the excess charge is dragged at 
 the same  velocity by the  flow.
 Correspondingly, an {\it  effective} excess charge density
 $\bar{Q}_{v}^{\text{eff}}$
 smaller than the total excess charge density $\bar{Q}_{v}$
has to be employed \citep[e.g.,][]{jougnot2012derivation, revil2013coupled}. In  the  considered 1D case, the  source current density 
takes  the  form \citep[e.g.,][]{jardani_et_al_2010,jougnot2013seismoelectric}
\begin{equation}\label{js_1D}
j_s=\bar{Q}_{v}^{\text{eff}} i\omega w, 
\end{equation}
where $i\omega w$ is the relative fluid velocity. 
The effective excess charge formulation, which
 allows us to explicitly express the role played by the relative fluid velocity 
in the source current density generation,
is equivalent to the electrokinetic 
coupling coefficient formulation commonly used in the seismoelectric literature \cite[e.g.,][]{pride94}. 
The relationship between the effective excess charge and the electrokinetic coupling coefficient can be 
found in many works such as \cite{revil2004constitutive}, \cite{jougnot2012derivation}, or \cite{revil2013coupled}.

In  the  absence  of  an  external  current  density,  the 1D electrical
potential $\varphi$ in  response to a given source  current density satisfies
\citep[][]{Sill1983SP}
\begin{equation}\label{Sill_1D}
\frac{\partial}{\partial z} \left( \sigma \frac{\partial \varphi}{\partial z}
\right)= \frac{\partial j_{s}}{\partial z},
\end{equation}
where  $\sigma$  denotes the   electrical conductivity,  which
strongly depends  on the saturating  fluid as well as on textural  properties of the medium, such as porosity and tortuosity.

 The approach presented above is based on the fact that the
electrical potential field has a negligible effect on the fluid
flow pattern and, therefore, the seismic and electrical problems can be assumed to be
decoupled. This assumption is made in most seismoelectric studies for
materials similar to the ones considered in this work
\cite[e.g.,][]{haines2006seismoelectric, guan2008finite, zyserman2010finite}.

In order to obtain the  seismoelectric response of the sample shown in
Fig.~\ref{three_layers}, we must
solve  Eqs. (\ref{js_1D})  and (\ref{Sill_1D})  for  the relative
fluid   displacement   given   by   Eq.~(\ref{w_solution}). The general
solution for the electrical potential $\varphi$ is then given by
\begin{equation}\label{phi}
\varphi(z)=
\begin{cases}
\begin{array}{lcl}
-\frac{i\omega \bar{Q}_{v,2}^{\text{eff}} }{\sigma_{2}} \frac{A_{2}}{k_{2}}
\left( e^{k_{2}z}+e^{-k_{2}(L+z)} \right)+R^{-}_{2}z + S^{-}_{2},  \  \quad  
\quad  -L/2 \leq z \leq -L_{1},\\
-\frac{i\omega \bar{Q}_{v,1}^{\text{eff}} }{\sigma_{1}} \frac{A_{1}}{k_{1}}
\left( e^{-k_1 z}+e^{k_1 z} \right)+R_{1}z + S_{1}, 
\quad  \quad  \quad \quad  \quad   -L_{1}    \leq     z     \leq L_{1},\\
-\frac{i\omega \bar{Q}_{v,2}^{\text{eff}} }{\sigma_{2}} \frac{A_{2}}{k_{2}}
\left( e^{-k_{2}z}+e^{-k_{2}(L-z)} \right)+R^{+}_{2}z + S^{+}_{2},  \quad  \quad  
\quad  L_{1} \leq z \leq L/2.
\end{array}
\end{cases}
\end{equation}
The parameters $R^{-}_{2}$, $S^{-}_{2}$, $R_{1}$, $S_{1}$, $R^{+}_{2}$
and $S^{+}_{2}$  are complex-valued constants that can  be obtained by
imposing the following boundary conditions
\begin{eqnarray}
\frac{\partial \varphi}{\partial z}&=&0, \quad z=-L/2,L/2, \label{bc_phi_1}\\
\varphi&=&0, \quad z=L/2. \label{bc_phi_2}
\end{eqnarray}
Equation  (\ref{bc_phi_1})  states that  the  rock  sample is  electrically
insulated   at   its   bottom   and  top   boundaries,   whereas
Eq.~(\ref{bc_phi_2}) indicates that the top boundary  is the zero reference for the
electrical potential.
%From  Eq. (\ref{bc_phi_1})  it is clear  that  both
%$R^{+}_{2}$ and $R^{-}_{2}$  must be zeros.  

To  obtain an  additional condition, we integrate Eq.~(\ref{Sill_1D})
in the upper half of the sample
\begin{equation}\label{integro}
\left. \sigma  \frac{\partial \varphi}{\partial z}\right |_0^{L/2}
= \left. j_s\right|_0^{L/2}.
\end{equation}
Since $w=0$ in both the top boundary and the center of the sample, the
right-hand side of Eq.~(\ref{integro}) is zero. Using  Eq.~(\ref{bc_phi_1}) then gives
\begin{equation}\label{bc_phi_3}
\frac{\partial \varphi}{\partial z}=0, \quad z=0.
\end{equation}
The  four   boundary  conditions  stated   in  Eqs.  (\ref{bc_phi_1}),
(\ref{bc_phi_2}), and  (\ref{bc_phi_3}), together with  the continuity
of $\varphi(z)$ at $z=\pm L_1$, provide six relations that allow us to
determine the complex-valued constants of Eq.~(\ref{phi}).
By solving this linear system of equations, we obtain
\begin{equation}\label{phi_final}
\varphi(z)=
\begin{cases}
\begin{array}{lcl}
-\frac{i\omega \bar{Q}_{v,1}^{\text{eff}} }{\sigma_{1}} \frac{A_{1}}{k_{1}}
\left( e^{-k_1 \lvert z\rvert}+e^{k_1 \lvert z\rvert} \right)+ S_{1}, 
\quad  \quad  \quad \quad  \quad \quad     0 \leq  \lvert z\rvert \leq L_{1},\\
-\frac{i\omega \bar{Q}_{v,2}^{\text{eff}} }{\sigma_{2}} \frac{A_{2}}{k_{2}}
\left( e^{-k_{2}\lvert z\rvert}+e^{-k_{2}(L-\lvert z\rvert)} \right)+ S_{2},  \quad  \quad  
\quad  L_{1} \leq \lvert z\rvert \leq L/2,
\end{array}
\end{cases}
\end{equation}
where $S_{1}$ and $S_{2}$ are given by
\begin{equation}
S_{1}=\frac{i\omega                          \bar{Q}_{v,1}^{\text{eff}}
}{\sigma_1}\frac{A_{1}}{  k_{1}} \left( e^{-k_{1}L_{1}}+e^{k_{1}L_{1}}
\right)-    \frac{i\omega   \bar{Q}_{v,2}^{\text{eff}}}{\sigma_{2}   }
\frac{A_{2}}{k_{2}}    \left(    e^{-k_{2}L_{1}}+e^{-k_{2}(L-L_{1})}-2
e^{-k_{2}L/2}\right),
\label{S_1}
\end{equation}
\begin{equation}
S_{2}=\frac{2i\omega                        \bar{Q}_{v,2}^{\text{eff}}}
{\sigma_2}\frac{A_{2}}{ k_{2}} e^{-k_{2}L/2}.
\label{S_2}
\end{equation}
Equation (\ref{phi_final}), together with Eqs.~(\ref{S_1}) and
(\ref{S_2}),  constitute  the   analytical solution  describing  the
seismoelectric  response  of a  rock  sample  containing a central horizontal
layer that is  subjected to
an oscillatory compressibility test (Fig.~\ref{three_layers}).

%%%%%%%%%%%%%%%%%%%%%%%%%%%%%%%%%%%%%%%%%%%%%%%%%%%%%%%%%%%%%%%%%%%%%%
\subsection{Particular solution: seismoelectric response of a rock sample containing a horizontal fracture}
Fractures are present in virtually all geological formations, and they
tend to  control the overall hydraulic  and mechanical properties of  these formations.
  This is  why there  is great interest  in developing
techniques to characterize fractured materials. Due to the very strong
compressibility  contrast typically observed between  fractures and the host rock,
wave-induced fluid flow and, therefore,  the corresponding  seismoelectric  effects, are
expected   to    be   significant   in   these environments.   Indeed,
\citet{jougnot2013seismoelectric}   showed   that   
 under typical laboratory  conditions the  presence   of
fractures can produce  measurable seismoelectric signals in response
to the application of an oscillatory compressibility test.  Here,  we
adapt the general analytical solution derived above 
to the particular case of a rock sample containing a horizontal fracture
at its center.

The poroelastic  response of a fractured  rock can be  obtained in the
framework  of  Biot's  (\citeyear{biot62})  theory by  representing  the
fractures as  highly compliant and permeable  heterogeneities embedded
in  a   stiffer  and  less   permeable  host rock.  In   this  sense,
\citet{brajanovski_et_al05} employed an  analytical solution of Biot's
(\citeyear{biot62}) equations for periodically varying coefficients to
compute seismic  attenuation and dispersion  due to wave-induced fluid flow in fractured
rocks. They considered the  limit of very small values for
the stiffnesses and apertures of the fractures in conjunction with very
high  porosity, which allowed  for obtaining  
simple  expressions  for  the  corresponding effective  complex-valued
frequency-dependent plane wave modulus.

Following the ideas of \citet{brajanovski_et_al05},
 we  propose an analytical approach to
obtain  the seismoelectric  response  of a  rock  sample containing  a
horizontal fracture at its center.   For this purpose, the  analytical
solution obtained for a horizontal layer can be appropriately adapted by 
considering  in Eq.  (\ref{phi_final}) that the aperture  and porosity of  
the fracture satisfy $h=2L_1  \rightarrow  0$ and  $\phi^f\rightarrow  1$.
Under   these  assumptions,  the
contribution of the  fracture to wave-induced fluid flow can be significant only  if 
the  fracture stiffness also  becomes very small.  To  take
into account this inter-dependence, and following \citet{brajanovski_et_al05}, we
characterize the  elastic properties of the drained  fracture in terms
of  the shear compliance  $Z_T$ and  drained normal  compliance $Z_N$,
which are given by
\begin{eqnarray}
Z_{T}&=&\lim_{h \to 0} \frac{h}{\mu_m^f},\\
Z_{N}&=&\lim_{h \to 0} \frac{h}{K_m^f+\frac{4}{3}\mu_m^f},\label{zn}
\end{eqnarray}
where  $K_{m}^{f}$ and  $\mu_{m}^{f}$ are  the drained-frame  bulk and
shear moduli of the  fracture, respectively.  By taking the above
limits, and  using the fracture shear and  normal  compliances, Eq.
(\ref{phi_final}) takes the form
\begin{equation}\label{phi_NAKA}
\varphi(z)=  -\frac{i\omega  \bar{Q}_{v,2}^{\text{eff}}  }{\sigma_{2}}
\frac{\bar                    {A_{2}}}{k_{2}}                   \left(
e^{-k_{2}\lvert z\rvert}+e^{-k_{2}(L-\lvert z\rvert)}-2e^{-k_{2}(L/2)}\right),
\end{equation}
where
\begin{equation}\label{A_2_bar}
\bar {A_2}=\lim_{h\rightarrow 0} A_{2}=
\frac{\Delta P \left(1-\beta_2\right)}
{\frac{2}{Z_{N}}\left( 1-e^{-k_{2}L}\right)+
N_{2}k_{2}\left( 1+e^{-k_{2}L}\right)}.
\end{equation}
This analytical solution allows for computing the
seismoelectric  response  of a  rock  sample  containing a  horizontal
fracture  at its center. Note that the only fracture parameter involved
in these equations is the  drained normal compliance $Z_{N}$
and the only structural parameter is the total thickness of the sample $L$.

%%%%%%%%%%%%%%%%%%%%%%%%%%%%%%%%%%%%%%%%%%%%%%%%%%%%%%%%%%%%%%%%%%%%%%
\section{Results}
%%%%%%%%%%%%%%%%%%%%%%%%%%%%%%%%%%%%%%%%%%%%%%%%%%%%%%%%%%%%%%%%%%%%%%

In  this section, we  employ the general and particular analytical
solutions derived above to explore the roles played by different petrophysical and structural
properties of heterogeneous porous rocks on the seismoelectric signals
produced  by  oscillatory  compressibility  tests. In  all  cases,  we
assume that the pore fluid is water and that the solid grains consist
of quartz (see Table~\ref{materialprop}).

%%%%%%%%%%%%%%%%%%%%%%%%%%%%%%%%%%%%%%%%%%%%%%%%%%%%%%%%%%%%%%%%%%%%%%
\subsection{Petrophysical relationships}

For the analysis based on the general solution,
we consider a poroelastic model corresponding to a rectangular rock
sample containing a  horizontal layer at its  center (Fig.~\ref{three_layers}).
Both the layer as well as the host rock correspond to clean sandstones, albeit with different 
porosities. To  relate the porosity $\phi$ to the permeability $\kappa$,
we use the Kozeny-Carman equation 
 \citep[e.g.,][]{Mavko_et_al_2009}
\begin{equation}\label{Kozeny-Carman}
\kappa=b\frac{\phi^{3}}{\left(1-\phi\right)^{2}}d^{2},
\end{equation}
where $b$  is a geometrical factor  that depends on  the tortuosity of
the porous medium, and $d$  the mean grain diameter. In this analysis,
we take $b=0.003$ and $d=8\times 10^{-5}$ m. In addition to changes in
permeability, porosity variations also imply changes in the mechanical
properties.   To link  the porosity  and the
solid grain properties with the elastic moduli of the dry frame, we use
the empirical model of \citet{krief-et-al90}
\begin{equation}
K_{m}=K_{s}\left(1-\phi\right)^{3/\left(1-\phi\right)},\label{Krief_Km}
\end{equation}
\begin{equation}
\mu=\frac{ K_{m}\mu_{s}}{K_{s}},\label{Krief_mu}
\end{equation}
where $\mu_{s}$ is the shear modulus of the solid grains.

\begin{table}
\begin{center}
\caption{Material properties employed in the analysis considered in this work.} 
	\begin{tabular}{lccccc}
	\hline	
        \hline
	Quartz grain bulk modulus, $K_s$ [GPa]& $37$\\
 	Quartz grain shear modulus, $\mu_s$ [GPa] & $44$\\
	Water bulk modulus, $K_f$ [GPa]& $2.25$ \\
        Water viscosity, $\eta$ [$\text{Pa}\times{s}$] & $0.001$ \\
        Water electrical conductivity, $\sigma_w [\text{S} \ \text{m}^{-1}]$ & $0.01$\\
        Water density, $\rho_w [\text{Kg} \ \text{m}^{-3}]$ & $ 10^{3}$\\
        \hline
               & Material 1 & Material 2 & Material 3\\
        \hline
%	Grain bulk modulus, $K_s$ [GPa]& $37$ &  $37$ &   $37$ &  $37$  &  $37$\\
% 	Grain shear modulus, $\mu_s$ [GPa] & $44$ & $44$ & $44$ & $44$  & $44$\\
        Porosity, $\phi$ & $0.05$ & $0.2$ & $0.4$ \\
	Dry rock bulk modulus, $K_m$ [GPa]& $31.47$ & $16.02$ &  $2.88$\\
 	Dry rock shear modulus, $\mu$ [GPa] & $37.42$ & $19.05$ & $3.42$\\
	Permeability, $\kappa$ [mD] & $2.66$ & $240$ & $3410$\\
        Electrical conductivity, $\sigma \ [\text{S} \ \text{m}^{-1}]$ & $2.5\times 10^{-5}$ & 
        $4 \times 10^{-4}$ & $1.6 \times 10^{-3} $\\
        Effective excess charge density, $\bar{Q}_{v}^{\text{eff}} \ [\text{C} \ \text{m}^{-3}]$ 
        &$526.8$ & $13.13$ & $1.49 $ \\
        Biot's critical frequency, $f_{B}$ [Hz] & $2.99\times 
        10^{6}$ & $1.32\times 10^{5}$ & $1.8\times 10^{4}$ \\  
	\hline
        \hline
 	\end{tabular}

\label{materialprop}
\end{center}
\end{table}

The  electrical  conductivities   $\sigma$  of  the  considered  clean
sandstones are  obtained using the empirical  relationship proposed by
\cite{archie1942electrical}
\begin{equation}\label{Archie}
\sigma = \phi^m \sigma_w,
\end{equation}
where  $m$ = 2  is the  cementation exponent  and $\sigma_w$  the   
electrical conductivity of the pore water.   The remaining  electrical parameter,
$\bar{Q}_{v}^{\text{eff}}$,  is obtained  by  employing the  empirical
relationship proposed by \cite{jardani2007tomography}
\begin{equation}\label{Jardani_empirical}
\log \left(\bar{Q}_{v}^{\text{eff}}\right)=-9.2349 -0.8219 \log(\kappa),
\end{equation}
where $\kappa$ and $\bar{Q}_{v}^{\text{eff}}$ are in units of m$^2$ and
C/m$^3$, respectively. Below Biot's critical frequency the
effective excess  charge density is  similar to the one at
zero frequency  \citep[e.g.,][]{tardif2011frequency, revil2013coupled} and,
hence, boundary layer effects can be neglected in the test cases considered in the following.
Even if we consider simple petrophysical relationships to link $\sigma$ and $\bar{Q}_{v}^{\text{eff}}$
to porosity, these properties can also be determined independently 
in laboratory experiments
\citep[e.g.,][]{jouniaux1995streaming,suski2006monitoring}.

For the analysis of the particular solution, we consider a rectangular homogeneous sandstone
containing a horizontal fracture at its center. 
Equations (\ref{Kozeny-Carman}) to  (\ref{Jardani_empirical}) are employed to determine the petrophysical properties of the
host rock, while the parameter $Z_N$ characterizes those of the fracture.

%%%%%%%%%%%%%%%%%%%%%%%%%%%%%%%%%%%%%%%%%%%%%%%%%%%%%%%%%%%%%%%%%%%%%%
\subsection{General solution analysis}

\subsubsection{Compliant, high-permeability layer }

\begin{figure}
\begin{center}
  \includegraphics[angle=0,,width=1\textwidth]{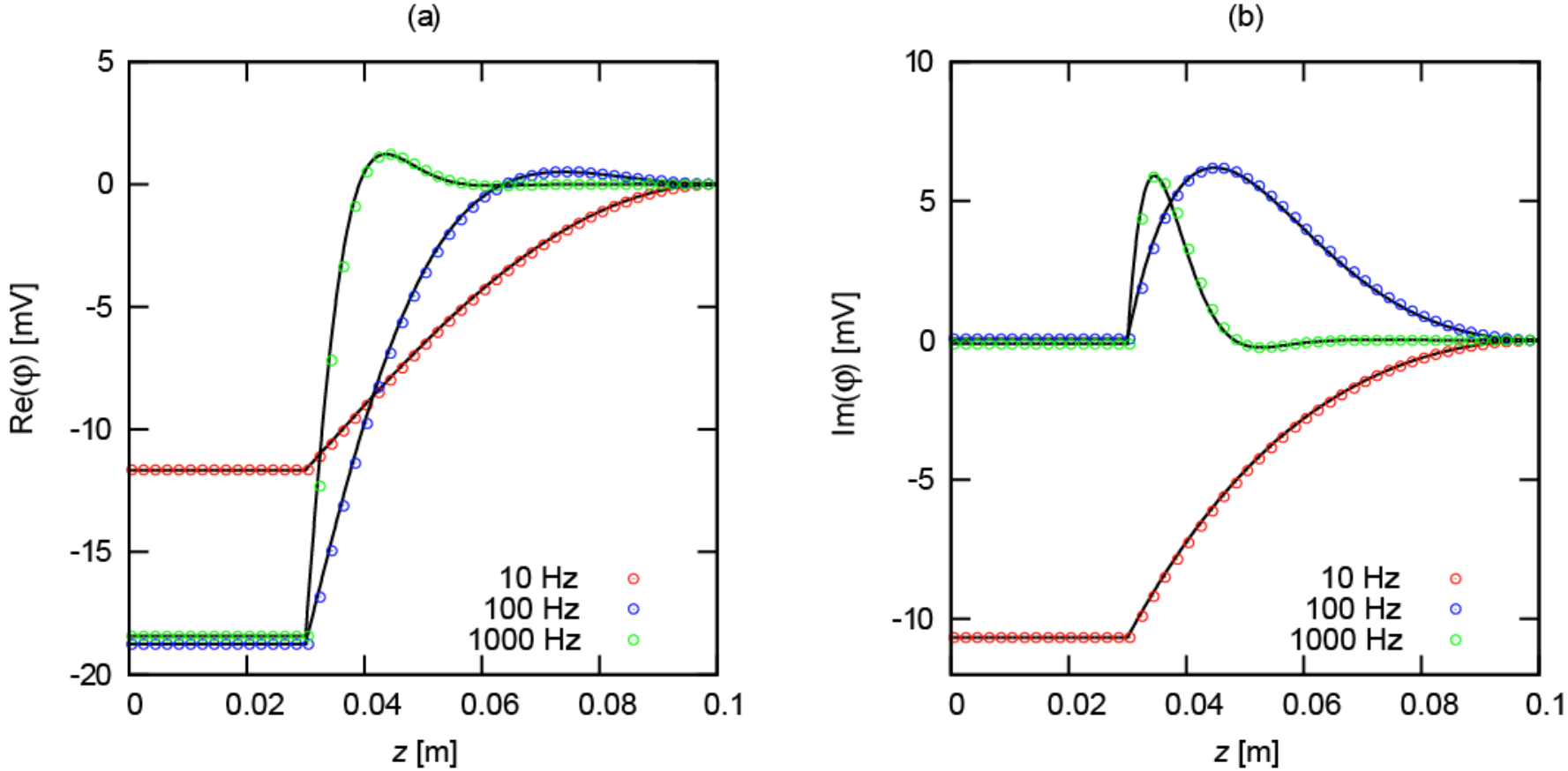}
\caption{Comparison between (a) the real and (b) the imaginary parts of the electrical potential
of a rock sample containing a horizontal layer at its center based on the 
analytical solution derived in this study (solid lines) and the numerical framework by 
\cite{jougnot2013seismoelectric} (circles) as functions of position. The different
curves correspond to three different frequencies of the applied harmonic compression.
The imaginary part is related to the  lag between the applied oscillatory compression and the resulting
electrical potential distribution.} 
\label{Fig_2_num}
\end{center}
\end{figure}

We first consider a rock  sample with a vertical sidelength  of 20~cm composed
of a  stiff, low-permeability host rock with a porosity of  0.05 (material 1 in Table~\ref{materialprop}),
 permeated at its  center by a
compliant, high-permeability horizontal layer with a thickness of 6~cm and a
porosity  of 0.4 (material 3 in Table~\ref{materialprop}).    The sample
is  subjected to  a harmonic  compression  of amplitude  $\Delta P=  1
\  \text{kPa}$  and frequencies  ranging from 1 to 10$^4$  Hz.
Before analyzing the seismoelectric response of this sample, we show in 
Fig.~\ref{Fig_2_num} a comparison between the electrical potential 
calculated with the analytical solution described in this work and the
solution obtained using
the numerical framework proposed by \cite{jougnot2013seismoelectric}. Due to the
symmetry of the solution, we only show the response in the upper half of the sample.
We observe excellent agreement between the electrical potential curves obtained
using the analytical and numerical methodologies.

\begin{figure}
\begin{center}
  \includegraphics[angle=0,,width=1\textwidth]{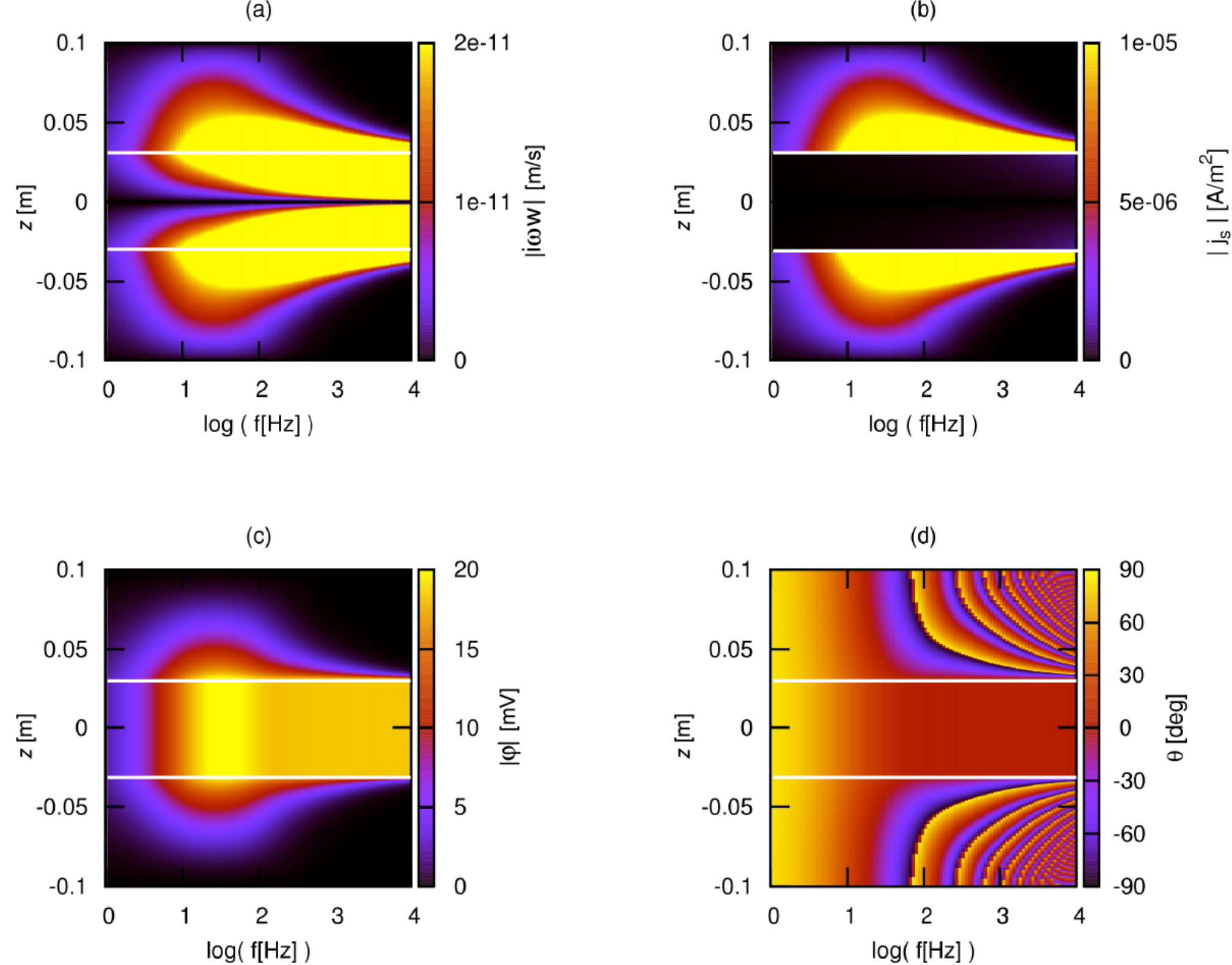}
\caption{(a) Relative fluid velocity, (b) electrical source current density,
 and (c) amplitude and (d) phase of the electrical potential corresponding to a rectangular,
 stiff, low-permeability host rock containing a compliant, high-permeability horizontal layer at
  its  center. The porosity
  of the layer is 0.4 (material 3 in Table~\ref{materialprop}), whereas that of the host rock
  is 0.05 (material 1 in Table~\ref{materialprop}). 
 In all cases, the panels
  show the parameters along the $z$-axis as functions of
  the frequency of the applied harmonic compression. For visualization purposes, we indicate
the boundaries of  the layer using white  lines.}
\label{Fig_2_top}
\end{center}
\end{figure}

Returning to the analytical solution for the compliant layer, 
Fig.~\ref{Fig_2_top}(a) shows the  amplitude of the resulting relative
fluid  velocity along  the  $z$-axis of the sample as  a function  of
frequency. 
Due to the strong  contrast between the compressibilities of the
two materials, significant relative fluid velocities arise in both the
host rock  and  the layer. 
There is a frequency at which the spatial extension 
of host rock affected by fluid flow is at its maximum, which we denote as the
peak frequency. The characteristic
frequency  $f_c$ given  by Eq. (\ref{w_c}) is indicative of this frequency,
since for such a value the diffusion length in  the host rock is
comparable to the characteristic size of the region.
For lower frequencies, fluid velocities tend to  become negligible.
 For higher frequencies, the regions of the host rock affected by fluid
flow tend to only include the immediate vicinity  of the boundaries
of  the layer, but the
magnitude of the  relative fluid velocity is important and increases 
with frequency. Relative fluid velocity is also important
inside  the  layer, mainly  in the  vicinity  of  the
boundaries.  Since  the permeability of this material is much higher,
the corresponding peak frequency 
occurs at higher frequencies, as suggested by Eq. (\ref{diffusivity})
and \eqref{w_c}.

A significant  current density (Fig.~\ref{Fig_2_top}(b)) prevails 
in the  host rock due to  the relative fluid velocity field (Fig.~\ref{Fig_2_top}(a)) produced 
by  the compression.  Moreover, the source current density clearly depends on the frequency,
a relation that arises from the frequency-dependence of the induced fluid flow.
The maximum current densities occur at the contacts between the two materials,
where the relative fluid velocity is also maximum. Within the layer, even though
significant fluid flow also takes place, the  resulting source
current density is very small, since the effective 
excess charge is much smaller in this material characterized by a 
much higher permeability 
(Table~ \ref{materialprop}, Eq.~(\ref{Jardani_empirical})). 

Significant electrical potential differences (Fig.~\ref{Fig_2_top}(c)), well above the $\sim$ 0.01~mV threshold of laboratory experiments
\cite[e.g.,][]{zhu2005seismoelectric,schakel2012seismoelectric},
arise in response to the oscillatory compression. These results are consistent with
those of \cite{jougnot2013seismoelectric} for  fractured rocks
and points to the importance of wave-induced fluid flow effects on seismoelectric signals
 in  the presence  of porosity  variations.  In the host rock,
the  frequency-dependence  of  the  electrical
potential  distribution  is  very  similar  to  that  of  the  current
density. Moreover, the region in the host rock characterized by 
significant values of electrical potential has its
maximum  spatial extension   at the corresponding peak frequency.
Inside  the layer  the  behaviour  is
different. For each frequency, the amplitude of the electrical
potential is rather constant.  
This happens because in this high-permeability material
the  effective excess  charge is small 
while the electrical conductivity is  large, which implies that
the electrical potential is reduced
to a near-constant value,  $S_{1}$ (Eq.  (\ref{phi_final})).  Because  the electrical  potential is
continuous, this  constant corresponds to the value  of the electrical
potential at  the contact between  the two materials.   

The resulting  electrical potential is  not only characterized  by its
amplitude but  also by  its phase. For relatively low frequencies, the phase  $\theta$ of the electrical potential remains constant, while
for high frequencies it shows  rapid  changes within  the
host rock (Fig.~\ref{Fig_2_top}(d)).  Inside  the  layer $\theta$
remains constant, in  agreement with the behaviour observed for
the amplitude of the electrical potential in this region (Fig.~\ref{Fig_2_top}(c)).

%%%%%%%%%%%%%%%%%%%%%%%%%%%%%%%%%%%%%%%%%%%%%%%%%%%%%%%%%%%%%%%%%%%%%%
\begin{figure}
\begin{center}
  \includegraphics[angle=0,,width=1\textwidth]{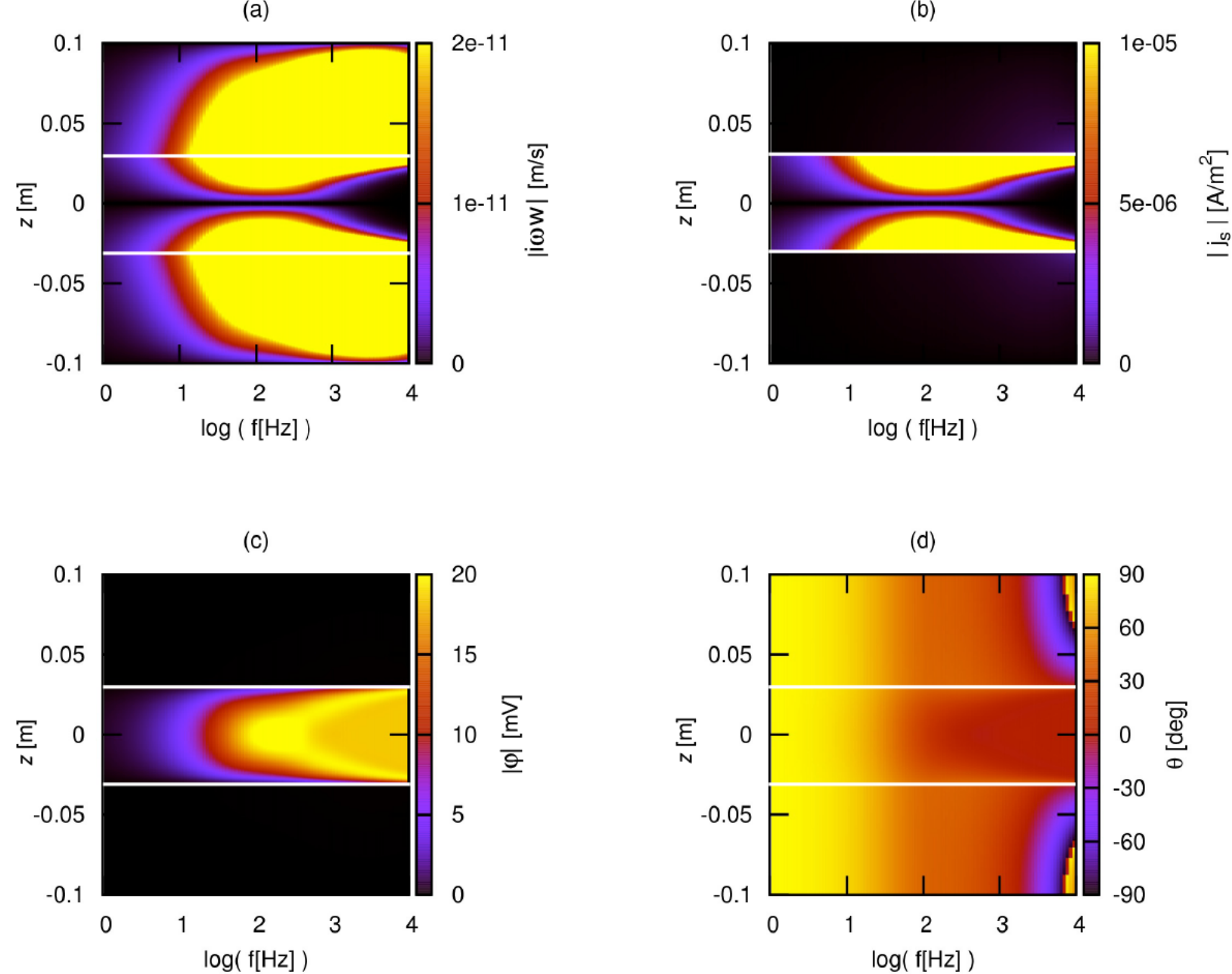}
\caption{(a) Relative fluid velocity, (b) electrical source current density,
 and (c) amplitude and (d) phase of the electrical potential corresponding to a rectangular,
 compliant, high-permeability host rock containing a stiff, low-permeability horizontal  layer at
  its  center.  The porosity
  of the layer is 0.05 (material 1 of Table~\ref{materialprop}), whereas that of the host rock 
  is 0.4 (material 3 in Table~\ref{materialprop}). 
  In all  cases, the panels
  show the parameters along the $z$-axis and as functions of
  the frequency of the applied harmonic compression. For visualization purposes, we indicate
the boundaries of  the layer using white  lines.}
\label{Fig_2_bot}
\end{center}
\end{figure}

\subsubsection{Stiff, low-permeability layer }

We now repeat the preceding analysis, but with the properties of the host
rock and of the layer interchanged. That is, we assume that the layer and
the host rock have porosities of  0.05 (material 1) and  0.4 (material 3), respectively, with the corresponding 
petrophysical properties given in Table~\ref{materialprop}.
Due to the strong compressibility contrast between
the two materials, significant fluid flow arises in the vicinity of the contact zones between 
the host rock and the layer (Fig.~\ref{Fig_2_bot}(a)).
Moreover, we observe that the peak frequency 
corresponding to the stiff, low-permeability material is 
higher compared with the previous situation (Fig.~\ref{Fig_2_top}(a)). The reason for this is that the
thickness of the stiff material is now smaller and, therefore, as dictated by Eq. \eqref{w_c},
the characteristic frequency shifts towards higher frequencies compared to the situation depicted
in Fig.~\ref{Fig_2_top}(a).
We also observe that for lower frequencies the fluid velocities tend to be negligible.
For higher frequencies, significant fluid velocities prevail in smaller spatial extensions of the layer but their
magnitudes increase with frequency. In the compliant host rock,
 the peak frequency is higher because this material is much more permeable (Eq. (\ref{w_c})), which  explains  the  differing
behaviour  of the fluid  velocity as  compared with  that observed  in the
stiff material.

The highest magnitudes of the electrical source current density are now concentrated inside 
the low permeability layer (Fig.~\ref{Fig_2_bot}(b)), which is characterized by high effective excess charges. 
As a result, and due to the imposed boundary conditions, when the layer is much stiffer
and less permeable than the host rock, the electrical potential has a significant 
amplitude only inside the layer (Fig.~\ref{Fig_2_bot}(c)). 
Moreover, the electrical potential amplitude is frequency dependent, 
with a maximum at the centre of the layer and for a frequency close to the peak frequency 
(Fig.~\ref{Fig_2_bot}(a)). With respect to the phase of the electrical
potential, significant
phase  changes  within  the  sample  take  place  at  very  high
frequencies and in the host rock (Fig.~\ref{Fig_2_bot}(d)).

%%%%%%%%%%%%%%%%%%%%%%%%%%%%%%%%%%%%%%%%%%%%%%%%%%%%%%%%%%%%%%%%%%%%%%

\subsubsection{Sensitivity to the thickness of the layer}

%%% Explanation Figure 5

\begin{figure}
\begin{center}
  \includegraphics[angle=0,width=1\textwidth]{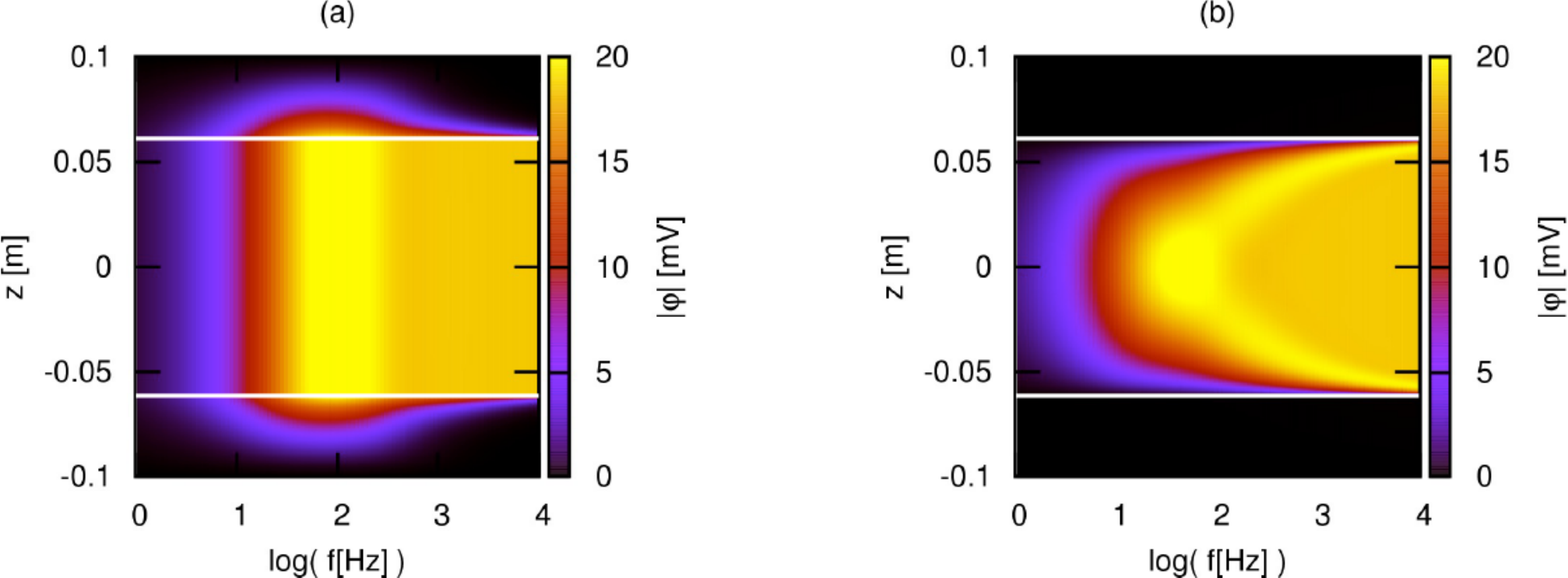}
\caption{Electrical   potential amplitude $\vert\varphi\vert$   for  a
  layer  having a thickness of $2  L_1 = $ 12  cm, that is,
  twice  the   value  considered  in   the  cases  shown   in  Figs.
  \ref{Fig_2_top}  and  \ref{Fig_2_bot}.   Panel  (a)  corresponds  to
  a compliant, high-permeability layer (material 3 in Table~\ref{materialprop}) 
embedded in a stiff, low-permeability host rock  (material 1 in Table~\ref{materialprop}),   
 whereas   panel  (b)  shows  the  results
  obtained when the two materials are interchanged. For visualization purposes, we indicate the boundaries
of the layer using white lines.}
\label{Fig_4}
\end{center}
\end{figure}

To analyze the role played by the thickness  of the layer, we
now consider $2 L_1=$ 12 cm for this region, that  is, twice the original value,
but leave the overall model setup and total thickness of the sample otherwise unchanged.
The frequency- and space-dependence as well as the magnitude of the 
electrical potential for the compliant and stiff
layer models, shown in Figs.~\ref{Fig_4}(a) and ~\ref{Fig_4}(b), 
are similar to the ones shown in Figs.~\ref{Fig_2_top}(c) and ~\ref{Fig_2_bot}(c), respectively. However, the maxima of the electrical
potential do  not occur at the same frequencies.  This shift in frequency can be explained
by the dependence of the characteristic frequency $f_c$, which indicates
the frequency at which maximum penetration of fluid flow takes place, on the characteristic
size of the involved medium. Indeed, Eq.~(\ref{w_c}) dictates that 
$f_c$ is inversely proportional to  the characteristic size of the region where fluid
flow occurs. When the layer  is more compliant and permeable
than the host rock,
the  electrical potential  is mainly  produced  by fluid  flow in  the
host rock.  As the thickness  of  the layer increases with respect to
the situation depicted in Fig.~\ref{Fig_2_top}(c),  the
characteristic  size of  the host rock  is reduced. This reduction, in turn,
produces an increase of  the characteristic frequency $f_c$, which can
be verified by comparing the frequency ranges where maximum electrical
potential  occurs in Fig.~\ref{Fig_2_top}(c)  and Fig.~\ref{Fig_4}(a).
Conversely,  when  the  layer  is  stiffer and less permeable than 
the  host rock, the electrical potential is produced by fluid flow inside the layer.
Therefore, as the thickness  of this region increases in the situation shown in 
Fig.~\ref{Fig_4}(b), the  characteristic frequency
$f_c$    is   reduced,   in    agreement   with    Eq.    (\ref{w_c}).
Correspondingly, the frequency range where the electrical potential is
maximum shifts towards lower  frequencies, as can  be verified 
by comparing Fig.~\ref{Fig_2_bot}(c)
and  Fig.~\ref{Fig_4}(b).   This  result therefore indicates that  the
seismoelectric signal  is sensitive to the  layer thickness, which largely controls 
the frequency range where the maximum signal occurs.

%%%%%%%%%%%%%%%%%%%%%%%%%%%%%%%%%%%%%%%%%%%%%%%%%%%%%%%%%%%%%%%%%%%%%%

\subsubsection{Sensitivity to the compressibility contrast between layer and host rock}

\begin{figure}
\begin{center}
  \includegraphics[angle=0,,width=1\textwidth]{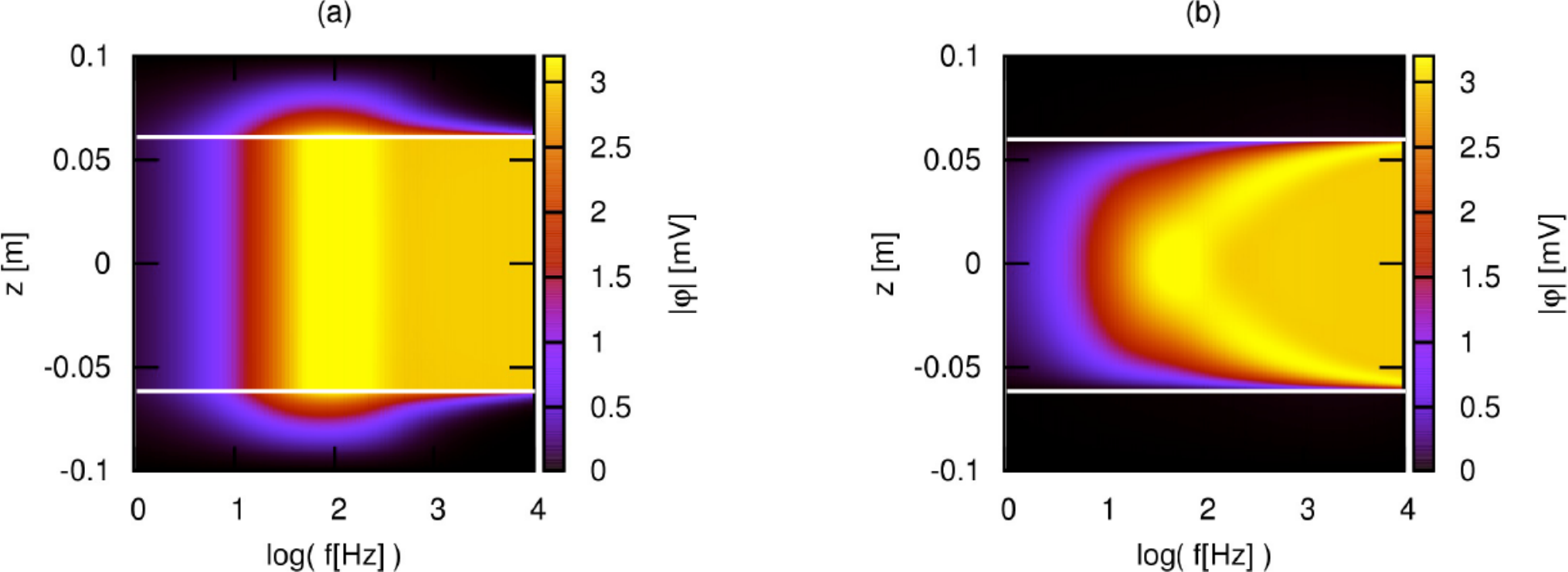}
\caption{Electrical potential amplitude $\vert\varphi\vert$ for a rock
  sample containing  a horizontal layer  of thickness $2 L_1  = 12~$cm
  and characterized  by a reduced  porosity contrast as  compared with
  the situation  depicted in Fig. \ref{Fig_4}.   Panel (a) corresponds
  to  porosity  values of  0.2 (material 2 in Table~\ref{materialprop})   and 0.05 
(material 1 in Table~\ref{materialprop})  for  the  layer  and
  host rock,  respectively,  whereas  panel (b)  shows  the
  results obtained when the two materials are interchanged.
  For visualization purposes, we indicate the boundaries of the layer using white lines.}
\label{Fig_5}
\end{center}
\end{figure}

Since  the  amount  of  fluid  flow scales  with  the  compressibility
contrast       between       background      and       heterogeneities
\citep[e.g.,][]{muller-et-al10},  the  porosity  contrast between  the
layer and  the host rock  is expected  to  have a  major
influence in  the resulting  seismoelectric response. To  analyze this,
we present in  Fig.~\ref{Fig_5} the amplitude of the electrical
potential as a function of frequency and space for the same geometries
used in the cases shown  in Fig.~\ref{Fig_4} but considering a reduced
porosity contrast.  We now use porosities of  0.05 
(material 1 in Table~\ref{materialprop}) and 0.2 (material 2 in Table~\ref{materialprop})  for the
stiff and compliant  materials, respectively.
By comparing Figs.~\ref{Fig_4} and Fig.~\ref{Fig_5}, we can verify
that the considered change in porosity contrast does not significantly
affect the general frequency-dependence and spatial
distribution of  the electrical potential. This is  because the geometry
as well as  the properties of the stiff material,  which is the region
where  most  of  the  current   density  arises, are similar.
Conversely, there is an
important decrease  of the amplitude  of the electrical  potential. This effect is expected, 
as a reduction of porosity contrast implies a decrease of compressibility
contrast. The latter causes a reduction of the wave-induced fluid flow
\citep[e.g.,][]{muller-et-al10}  and, thus, of the source  current
density.  These results illustrate that the magnitude of
the electrical potential may contain information on the
compressibility contrast between the layer and the host rock.

%%%%%%%%%%%%%%%%%%%%%%%%%%%%%%%%%%%%%%%%%%%%%%%%%%%%%%%%%%%%%%%%%%%%%%

\subsubsection{Asymptotic analysis}

The previous parametric analysis suggests that key mechanical, structural and hydraulic properties
of the central layer and of the  host rock affect the seismoelectric response.
 In particular, while the compressibility contrast between the materials controls the
magnitude of the resulting seismoelectric signal, the thickness and the permeability of the less permeable region
determine the frequency range at which the response takes its maximum value. 

In order to provide a formal description of this dependence and to further explore the analytical expression
given by Eq. \eqref{phi_final}, we develop the low- and high-frequency asymptotic behaviors of the solution. 
Since from the previous analysis we conclude that, for a given frequency, the electrical
potential is at its maximum at $z=0$, we derive these limits at this particular place of the sample. 
In Appendix B, we show  that 
the low-frequency asymptote of the electrical potential at the center of the sample is given by
\begin{equation}\label{LF}
\varphi^{LF}_{z=0}(\omega)= \frac{i\omega \Delta P}{2} 
\left( \beta_1-
\beta_2\right) \frac{\sum_{j=1}^{2} 
\frac{\bar{Q}_{v,j}^{\text{eff}}L_{j}} {\sigma_{j}}}
{\sum_{j=1}^{2}\frac{N_j}{L_{j}}},
\end{equation}
whereas the high-frequency asymptote is
\begin{equation}\label{HF}
\varphi^{HF}_{z=0}(\omega) = -\Delta P \left(\beta_1-
\beta_2\right) \frac{\sum_{j=1}^{2}\frac{\bar{Q}_{v,j}^{\text{eff}}}
{\sigma_{j}}\sqrt{\kappa_{j}N_{j}}}
{\sum_{j=1}^{2}\sqrt{\frac{N_{j}}{\kappa_{j}}}}.
\end{equation}
The latter expression is the high-frequency limit of the solution
obtained based on the premise that frequency is lower than Biot's
critical frequency. In other words, Eq. \eqref{HF} represents the high-frequency limit of the low-frequency-domain solution.

\begin{figure}
\begin{center}
  \includegraphics[angle=0,,width=1\textwidth]{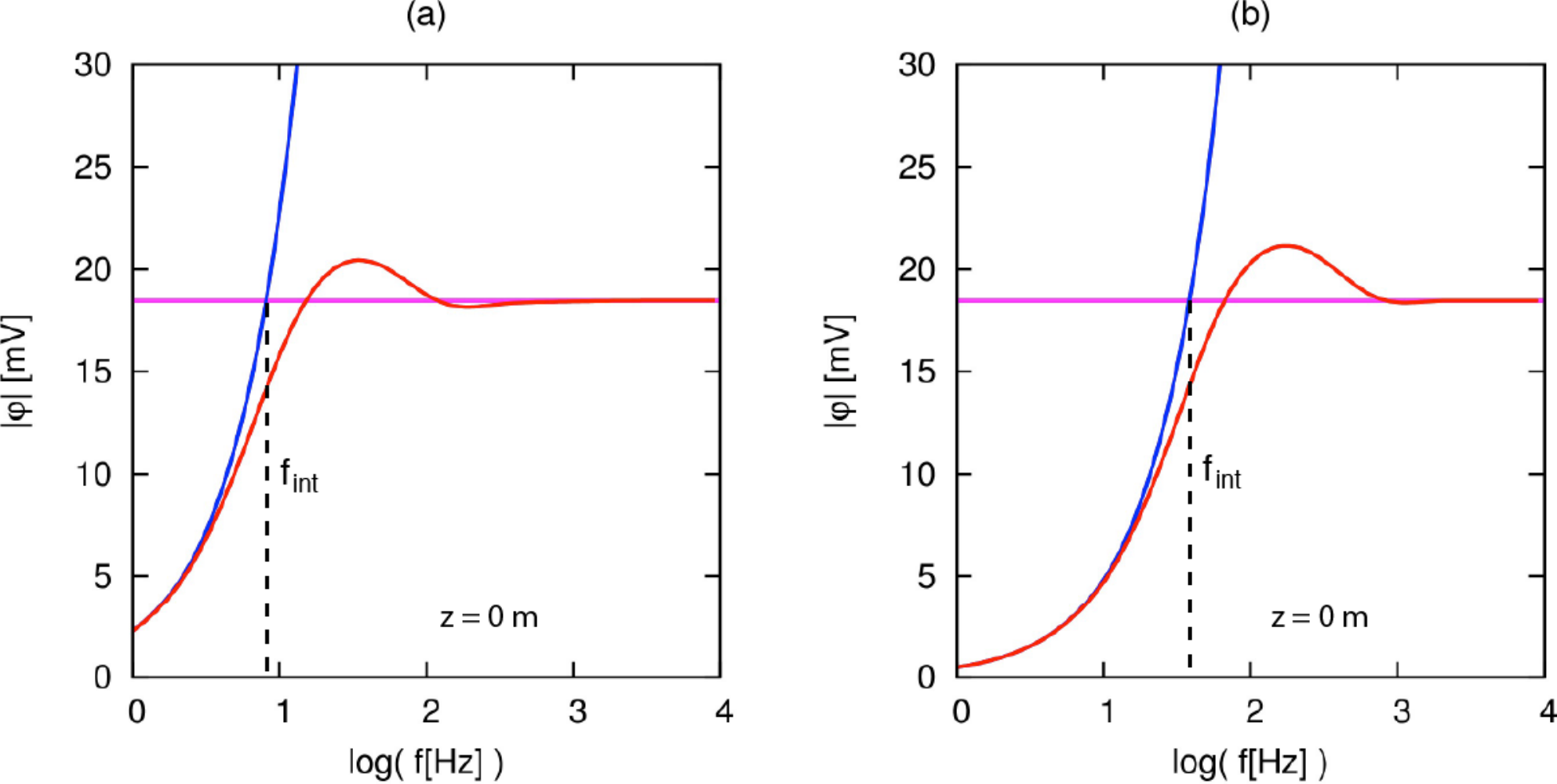}
\caption{Amplitude of the electrical potential in the center of the sample ($z=0$) as a function  
  of  frequency. Panel (a) corresponds to porosity values of
  0.4 (material 3 in Table~\ref{materialprop}) and 0.05 (material 1 in Table~\ref{materialprop})
 for the layer and host rock, respectively, whereas
  panel (b) shows the results obtained when the two materials are interchanged.
  Red lines correspond to the exact solution (Eq.  (\ref{phi_final})), the blue
  ones   to    their   low-frequency    asymptotic   solution   (Eq.
  \eqref{LF})  and  the  pink  ones to  their  high-frequency
  asymptotic solution (Eq. \eqref{HF}).}
\label{Plot_curvas}
\end{center}
\end{figure}

Figure~\ref{Plot_curvas} shows the amplitude of the electrical potential at the center
of the sample as a function of frequency, for both the compliant- and
stiff-layer cases considered 
in Figs.~\ref{Fig_2_top} and \ref{Fig_2_bot},
together with the corresponding low- and 
high-frequency asymptotes. There is very good agreement between the general solutions 
and the low- and high-frequency asymptotes. Moreover, this analysis
shows the importance of the analytical solutions obtained in this work and, in 
particular,  of these asymptotes, since they allow us to explicitly observe how the 
various petrophysical parameters 
contribute to the seismoelectric response. The high-frequency asymptote, which 
is constant, can be considered as an indicator of the maximum magnitude of the 
electrical potential. Therefore, Eq. \eqref{HF} allows for a direct appreciation of how the 
different  petrophysical properties control the  amplitude of the
seismoelectric response. This maximum amplitude is directly proportional to the applied 
stress, to a weighted average of the effective excess charges of the two media, and 
to the Skempton coefficient contrast between the host rock and the layer. 
The Skempton coefficient is the ratio of the fluid pressure increase to the corresponding
applied stress for undrained conditions  \citep[e.g.,][]{wang_book}. Correspondingly, the term
\begin{equation}
\Delta P \left(\beta_1-
\beta_2\right),
\end{equation}
appearing in Eq.  \eqref{HF} constitutes a measure of the  fluid pressure gradient induced across 
the interfaces separating the host rock and the layer. It is thus reasonable that the maximum amplitude of the
seismoelectric response turns out to be directly proportional to this parameter, since the
fluid pressure gradient induced by the applied compression triggers the fluid flow and the
corresponding seismoelectric signal.

Employing Eq. \eqref{Jardani_empirical}, the weighted average  of the effective excess charges included in Eq.  \eqref{HF}
satisfies
\begin{equation}\label{aprox1}
 \frac{\sum_{j=1}^{2}\frac{\bar{Q}_{v,j}^{\text{eff}}}
{\sigma_{j}}\sqrt{\kappa_{j}N_{j}}}
{\sum_{j=1}^{2}\sqrt{\frac{N_{j}}{\kappa_{j}}}} \propto  \frac{\sum_{j=1}^{2}\frac{\kappa_j^{-0.32}\sqrt{N_j}}
{\sigma_{j}}}
{\sum_{j=1}^{2}\sqrt{\frac{N_{j}}{\kappa_{j}}}}.
\end{equation}
If the permeability contrast between the two media is strong,
the numerator and denominator of the right-hand side of
Eq. \eqref{aprox1} are controlled by the term corresponding to the material of lower 
permeability, that is,
\begin{equation}\label{aprox2}
 \frac{\sum_{j=1}^{2}\frac{\bar{Q}_{v,j}^{\text{eff}}}
{\sigma_{j}}\sqrt{\kappa_{j}N_{j}}}
{\sum_{j=1}^{2}\sqrt{\frac{N_{j}}{\kappa_{j}}}} \propto  \frac{\kappa_{min}^{-0.32}\sqrt{N_{min}}}
{\sqrt{\frac{N_{min}}{\kappa_{min}}}}=\kappa_{min}^{0.18},
\end{equation}
where the subscript $min$ denotes the material with the lowest permeability. Substituting this relation
into Eq. \eqref{HF}, we obtain
\begin{equation}\label{HF_aprox}
\varphi^{HF}(\omega) \propto -\Delta P \left(\beta_1-
\beta_2\right)\kappa_{min}^{0.18}.
\end{equation}
Equation~\eqref{HF_aprox} suggests
that the magnitude of the resulting electrical potential, in the case of strong 
permeability contrasts, is directly proportional to $\kappa_{min}^{0.18}$. This 
result also indicates that the hydraulic properties of the most permeable material and 
the thicknesses of the layer and of the host rock 
do not affect the magnitude of the seismoelectric response. This 
is in agreement with the results presented in this work,
where similar values for the maximum amplitudes of the electrical potential were
found when the materials of the host rock and the layer were interchanged.

The frequency at which the low- and high-frequency asymptotes intercepts,
$f_{int}$, can be used as an
indicator of the frequency at which the electrical potential reaches its peak.
This intersection can differ with respect to the true peak frequency by almost
one order-of-magnitude  (Fig. \ref{Plot_curvas}). However, its dependence on the mechanical, hydraulic, and 
structural properties of the
heterogeneous sample are expected to also hold in the case of the true peak frequency. 
By requiring equality of Eqs. \eqref{LF} and \eqref{HF},
$f_{int}$ can be defined as
\begin{equation}\label{f_int}
f_{int}=\frac{1}{2\pi}\frac{\sum_{j=1}^{2}\frac{\bar{Q}_{v,j}^{\text{eff}}}{\sigma_{j}}\sqrt{\kappa_{j}N_{j}}}
  {\sum_{j=1}^{2}\sqrt{\frac{N_{j}}{\kappa_{j}}}}
  \frac{\sum_{j=1}^{2}\frac{N_j}{L_{j}}}                {\sum_{j=1}^{2}
    \frac{\bar{Q}_{v,j}^{\text{eff}}L_{j}} {\sigma_{j}}}.
\end{equation}
Hence, the frequency of the peak of the electrical potential does not depend on the 
amplitude of the applied pressure, $\Delta P$. Interestingly, it does not depend on the Skempton coefficient
contrast of the materials either.

Considering the case of strong permeability contrasts and using Eq. \eqref{HF_aprox} instead of Eq. \eqref{HF} 
to obtain $f_{int}$, yields
\begin{equation}\label{fint}
f_{int}\propto \frac{\kappa_{min}}{L_{min}}\sum_{j=1}^2\frac{N_j}{L_j}.
\end{equation}
This equation shows that the frequency of the peak of the electrical potential is directly proportional
to $\kappa_{min}$, whereas the hydraulic properties
of the most permeable regions do not affect this parameter.
In addition, the structural characteristics of the 
sample do affect the location of the  peak, which is 
inversely proportional 
to the thickness of the less permeable material.
Also, the mechanical properties, scaled by the thicknesses of the
corresponding layers, affect the peak frequency.

%%%%%%%%%%%%%%%%%%%%%%%%%%%%%%%%%%%%%%%%%%%%%%%%%%%%%%%%%%%%%%%%%%%%%%
\subsection{Analysis of fractured media}

Fractures are usually composed of  very compliant material and the resulting
strong  compressibility contrast  with  respect to  the host rock  is
expected to favor particularly strong seismoelectric
signals  in response  to the  application of  oscillatory compressions
\citep[][]{jougnot2013seismoelectric}. In order  to explore this
in more  detail, we estimate  the seismoelectric responses  of
fractured media by  making use of the model  developed in Section 2.3.
According  to this model,  in which  the aperture  of the  fracture is
considered  negligible, the seismoelectric  response is  controlled by
the  properties of  the host rock,  the size  of the  sample,  and the
compliance of the  fracture. The latter is represented  by the drained
normal compliance $Z_N$ (Eq. (\ref{zn})).

\begin{figure}
\begin{center}
  \includegraphics[angle=0,width=1\textwidth]{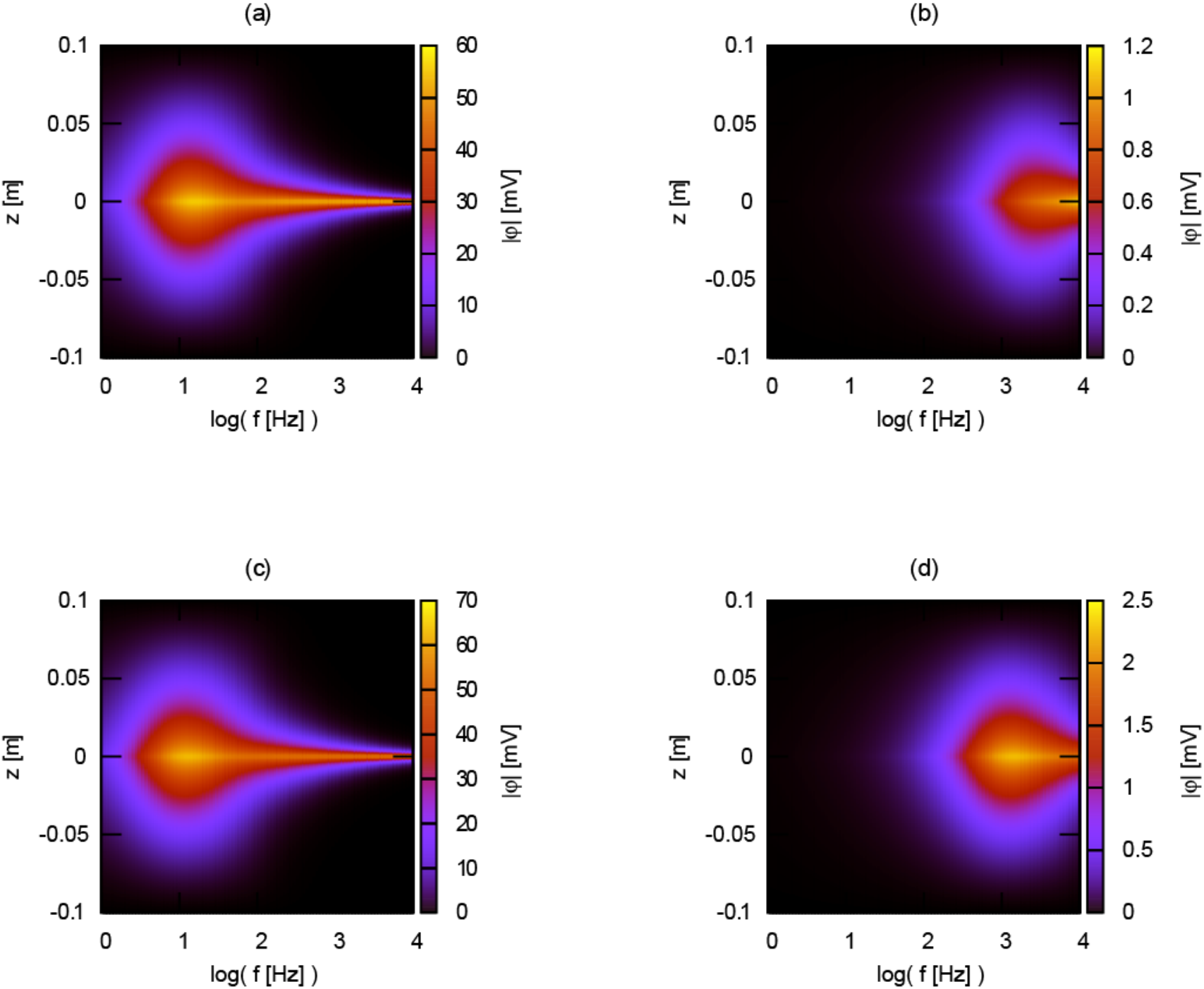}
\caption{Absolute     value     of     the    electrical     potential
  $\vert\varphi\vert$  along  the  $z$-axis  as  a  function  of
  frequency  for samples  containing  a horizontal  fracture at  their
  centers.   Panel (a)  corresponds to  a host rock with a porosity $\phi$ of
  $0.05$ (material 1 in Table~\ref{materialprop}) 
and a fracture drained normal compliance $Z_{N}$ of $10^{-8}
  \text{m  kPa}^{-1}$; in  panel  (b) $\phi=0.4$ (material 3 in Table~\ref{materialprop})  
  and $Z_{N}=  10^{-8}
  \text{m kPa}^{-1}$;  in panel  (c) $\phi=0.05$ and  $Z_{N}= 10^{-6}
  \text{m kPa}^{-1}$, and in  panel (d) $\phi=0.4$ and $Z_{N}= 10^{-6}
  \text{m kPa}^{-1}$.}
\label{Fig_6}
\end{center}
\end{figure}

To study the behaviour  of the seismoelectric response for different
characteristics of fracture and host rock materials, we consider a rectangular rock
sample with vertical sidelength of 20~cm containing  a horizontal fracture
at its center. As in the previous analysis, the sample is subjected to
a  harmonic  compression with an  amplitude  $\Delta  P=1$~kPa and  
frequencies ranging from 1 to 10$^4$~Hz. For the host rock, we consider a
sandstone    with a   porosity of   0.05 (material 1 in Table~\ref{materialprop}),
while,  as   proposed  by \citet{nakagawa-schoenberg07}, we use  a drained normal
compliance    $Z_{N}=10^{-8}~\text{m  kPa}^{-1}$ for the fracture.
Very  strong  electrical
potential  differences,  that  could  be measured using standard laboratory 
techniques, arise in response to  the applied oscillatory
compression (Fig.~\ref{Fig_6}(a)).  Moreover,  the overall frequency-  and space-dependence
characteristics of  the electrical potential are  essentially the same
as  in the case  of a  rock sample  containing a  horizontal compliant
and permeable layer.   Indeed, there is also a frequency range at which 
both the magnitude and the spatial extension of the electrical potential
are at a maximum.

To  study the  role  played by  the characteristics of the host rock,
we change its properties to those  of a  sandstone with a
porosity of 0.4 (material 3 in Table~\ref{materialprop}). The corresponding increases of
the permeability  and of the compressibility of the host rock  produce  significant
changes in the resulting electrical potential (Fig.~\ref{Fig_6}(b)). Indeed,  the  magnitude of  the signal  is
significantly smaller  than that  shown  in Fig.~\ref{Fig_6}(a), which
was evaluated using smaller values of permeability and
compressibility of the
host rock. This, in turn, indicates
that, according to Eq. \eqref{HF_aprox}, the effect related to the reduction of
compressibility contrast between fracture and host rock overcomes that related to the
increase of the permeability of the host rock. 
In addition,   as suggested by Eq. \eqref{fint}, the
increase  of the host rock  permeability shifts  the  frequency range 
where the maximum value and maximum spatial penetration  
of the  electrical  signal  occur  towards
higher values.

The  compliance of  the fracture,  represented by 
$Z_N$,  controls   the  compressibility  contrast  with regard to the 
host rock. This parameter is thus expected to
play  a  major  role  in  the  seismoelectric  response  of  fractured
materials. To  verify this, we show in  Fig.~\ref{Fig_6}(c) the electrical
potential for a fractured sample having a host rock porosity of 0.05,
that  is,  the same  value  employed  to calculate the electrical
potential represented in
Fig.~\ref{Fig_6}(a),   but    with   a   drained    normal   compliance
$Z_N=10^{-6}$~m kPa$^{-1}$. The comparison between Figs.~\ref{Fig_6}(a) 
and  ~\ref{Fig_6}(c) shows that the overall frequency-dependence and spatial 
distribution of the electrical potential are  very similar.
 However,  an increase  of fracture  drained normal
compliance  implies an  increase of  compressibility  contrast between
fracture and host rock.  This results in a significantly larger magnitude of the
electrical potential.

Finally, Fig.~\ref{Fig_6}(d) shows  the amplitude of the electrical potential
for a  host rock corresponding to a  sandstone with a  porosity of 0.4 and a
fracture with a drained normal compliance $Z_{N}= 10^{-6} \text{m kPa}^{-1}$.
By   comparing   this   electrical potential   with  the one
represented in 
Fig. ~\ref{Fig_6}(a),  we see a significant reduction  of the magnitude
of  the signal together  with a  shift towards  higher values  of the
peak frequency.   The  reduction in magnitude indicates  that  the increase  of
host rock  compressibility related  to  the porosity  change affects
the compressibility contrast more significantly than the increase of compliance of the fracture. 
In addition, as dictated by Eq. \eqref{fint}, the  increase  of  host rock permeability related  to  the  porosity
variation  shifts towards higher values the peak frequency of the electrical potential.

%%%%%%%%%%%%%%%%%%%%%%%%%%%%%%%%%%%%%%%%%%%%%%%%%%%%%%%%%%%%%%%%%%%%%%
\section{Discussion}
%%%%%%%%%%%%%%%%%%%%%%%%%%%%%%%%%%%%%%%%%%%%%%%%%%%%%%%%%%%%%%%%%%%%%%

In accordance with \cite{jougnot2013seismoelectric},
 our results indicate that seismoelectric signals are highly sensitive to
the presence of mesoscopic heterogeneities. The analytical solutions obtained
in this work provide the explicit dependence of these signals
on the structural, mechanical and hydraulic properties of the materials involved.
In the following, we discuss some of the implications of this dependence.

Assuming a laboratory experiment where electrodes are placed vertically along 
the sample shown in Fig.~\ref{three_layers}, our results suggest that one could identify 
the presence of a compliant and permeable layer as 
the region where the amplitude of the electrical potential remains constant,
and of a stiff and less permeable layer as the region exhibiting a significant electrical 
potential gradient. Alternatively, the center of the
layer can be identified 
by finding the position of the maximum 
amplitude of the potential. In the case of fractured rocks, the drained
normal compliance $Z_N$ is a parameter of key importance for simulating seismic wave 
propagation in the framework of the linear slip theory \citep[e.g.,][]{schoenberg80}, 
but it is often poorly constrained. Our 
analytical solution for the seismoelectric response of a fractured rock
could conceivably  be employed to estimate the drained normal compliance of 
a fracture based on electrical potential measurements. First, the position 
of the fracture could be identified as the region where the maximum amplitude is measured and, 
then, the drained normal compliance could be estimated by inserting the measured value of electrical potential 
in Eq.~(\ref{phi_NAKA}). Additionally, the permeability of the host rock 
could also be retrieved, provided that measurements are made at several distances 
from the fracture or at different frequencies.

The present analysis considers a rock sample subjected to an oscillatory compressibility test,
which allows us to explore seismoelectric signals related to the generation
of Biot's slow waves at interfaces separating layers with differing mechanical properties.
These results have interesting implications related to seismic wave propagation.
The seismoelectric signals are expected to decay as 
the amplitude of the seismic wave progressively diminishes 
due to geometrical divergence, scattering, and attenuation effects. 
Therefore, in the case of a compressional seismic wave propagating
through a periodic 1D medium composed of an assembly of the sample shown in
Fig.~\ref{three_layers}, similar to the situation studied by \citet{white-et-al}
in the context of seismic attenuation due to wave-induced fluid flow, the electrical-potential
contribution is expected
to be similar to that shown in this work, but scaled according
 to the local stress applied by the passing seismic wave.
A complete extension of the methodology presented in this work 
to seismic wave propagation is, however, not straightforward. 
In addition to scattering and amplitude distortion effects, 
further contributions,
such as the effects of wavelength-scale relative fluid flow produced by the seismic wave,
should be taken into account. The analysis of numerical simulations of seismic wave propagation
through porous media containing mesoscopic heterogeneities and of the resulting 
seismoelectric responses will be the subject of forthcoming studies.

Among the implications of the explicit dependence of the electrical potential
on rock parameters presented in this work, it is interesting to highlight that the magnitude
of the resulting signal is directly proportional to the Skempton coefficient 
contrast between the involved materials, to the amplitude of the applied compression and,
in the case of strong permeability contrasts, it is also highly influenced by the
permeability of the less permeable material. 
In such a situation the peak frequency of the electrical potential is largely
governed by the permeability and the thickness of the region containing the less permeable material.
Any attempt to model the coherent noise observed in seismoelectric 
signals arising potentially from  mesoscopic heterogeneities, as suggested by \cite{jougnot2013seismoelectric},
should carefully constrain the values of these parameters. 
Conversely,  in the case of strong permeability contrasts the hydraulic and electrical properties  
of the most permeable material do not need to be determined with high precision, as the seismoelectric 
signals turned out to be virtually insensitive  to these parameters.  

%%%%%%%%%%%%%%%%%%%%%%%%%%%%%%%%%%%%%%%%%%%%%%%%%%%%%%%%%%%%%%%%%%%%%%
\section{Conclusions}
%%%%%%%%%%%%%%%%%%%%%%%%%%%%%%%%%%%%%%%%%%%%%%%%%%%%%%%%%%%%%%%%%%%%%%

We have derived an analytical solution that describes the
seismoelectric response of a rectangular rock sample containing a horizontal
layer at its center that is subjected to an oscillatory compressibility experiment.  
The resulting solution was also adapted to compute the seismoelectric response
of a rock sample containing a horizontal fracture at its center.

The  seismoelectric responses predicted  by the analytical
solutions were explored and allowed us to shed some
light into their  dependence  on mechanical, hydraulic,
and  structural  properties of  the host rock and the layer or fracture. 
As a general result, we found that the seismoelectric signals produced by oscillatory
compressibility   tests  could  be    measured   in standard laboratory
experiments. The significant amplitudes of the seismoelectric signals 
point to the importance of considering mesoscopic
effects when employing the seismoelectic method. Moreover, the analysis
indicated  that  the maximum amplitude  of  the  electrical
potential  is directly proportional to the applied stress, to
the Skempton coefficient contrast  between  the host rock  and  the  heterogeneity,
and  to a weighted average of the effective excess charge of the two materials.
In presence of strong permeability variations we found that this weighted average is mainly controlled by the
permeability of the less permeable material and that the frequency at which the maximum
electrical potential prevails is  governed by the permeability and
thickness of the region containing such material. 

Although  the analytical solutions considered in this work are based  on  simple
1D  models, they constitute  useful tools  for exploring the
details of the physical processes involved in the seismoelectric method in 
presence of mesoscopic heterogeneities. Moreover, they may open the possibility
of retrieving key rock properties, such as permeability and fracture drained normal compliance,
from seismoelectric measurements.

\begin{acknowledgments}

This work  was supported by the Swiss National
Science    Foundation,   the    Agencia   Nacional    de   Promoci\'on
Cient\'{\i}fica y  Tecnol\'ogica (PICT 2010-2129),  Argentina, and the
Secretar\'{\i}a  de Ciencia  y T\'ecnica,  Universidad Nacional  de La
Plata, Argentina. L.B.M. gratefully acknowledges an extended visit to the University of Lausanne
financed by the Fondation Herbette. The authors also thank two anonymous reviewers and the Associated Editor for comments
and suggestions that helped to improve this paper.

\end{acknowledgments}

%%%%%%%%%%%%%%%%%%%%%%%%%%%%%%%%%%%%%%%%%%%%%%%%%%%%%%%%%%%%%%%%%%%%%%%%%%%
%%%%%%%%%%%%%%%%%%%%%%%%%%%%%%%%%%%%%%%%%%%%%%%%%%%%%%%%%%%%%%%%%%%%%%%%%%%

\appendix

\section{1D Skempton coefficient}
The Skempton coefficient is defined as the ratio between the induced fluid pressure increase to the applied
stress for undrained conditions \citep[e.g.,][]{wang_book}
\begin{equation}
\beta\equiv -\frac{\delta p_f}{\delta\tau}\Big\vert_{\zeta=0},
\end{equation}
where $\zeta$ is the increment of fluid content, which in 1D is given by  $\zeta\equiv -\partial w/\partial z$.
This parameter measures how the applied stress is distributed between the solid frame and the pore fluid
\citep[][]{wang_book}.

Imposing the undrained condition $\zeta=0$ in Eqs. \eqref{mod.2a} and \eqref{mod.2b}, we get
\begin{equation}\label{mod.2a.bis}
\tau=H\frac{\partial u}{\partial z},
\end{equation}
\begin{equation}\label{mod.2b.bis}p_f=  - \alpha  M \frac{\partial u}{\partial z}.
\end{equation}
The 1D Skempton coefficient is thus obtained by taking the ratio between Eqs. \eqref{mod.2a.bis} and \eqref{mod.2b.bis}
\begin{equation}
\beta=\frac{\alpha M}{H}.
\end{equation}

\section{Asymptotic analysis of the electrical potential at 
the center of the sample}

Equation~(\ref{phi_final}), together  with the Eqs. (\ref{S_1}) and
(\ref{S_2}),  constitute  an   analytical  model  for  describing  the
seismoelectric  response  of a  rock  sample  containing a  horizontal
layer that is subjected to an oscillatory compressibility test. Now we
focus on  the values of  $\varphi(z)$ at  $z=0$, that is, at the center 
of the sample. From Eq. (\ref{phi_final}), we get for the  electrical potential 
\begin{equation}\label{phi_cero_1}
\varphi(0)=
-2\frac{i\omega \bar{Q}_{v,1}^{\text{eff}} }{\sigma_{1}} \frac{A_{1}}{k_{1}}
+ S_{1}. 
\end{equation}
Replacing   the  corresponding   expressions  for   $A_1$   and  $S_1$
(Eqs. (\ref{A_1}) and  (\ref{S_1}) respectively) in (\ref{phi_cero_1})
we obtain
\begin{equation}\label{phi_cero_2}
\varphi(0)=\frac{i\omega \Delta P \left( \beta_1-
\beta_2\right)
 \sum_{j=1}^{2}\frac{\bar{Q}_{v,j}^{\text{eff}}} {\sigma_{j}k_{j}}
\left[ \frac{1-cosh(k_{j}L_{j})}{sinh(k_{j}L_{j})}\right]}
{\sum_{j=1}^{2} N_j k_j \ coth (k_jL_j)}.
\end{equation}
This equation predicts that the amplitude of the electrical potential
is proportional to the Skempton coefficient contrast of the two involved materials
and to the amplitude of the applied compression. 
However,      the     dependence      of     $\varphi      (0)$     on
$\bar{Q}_{v,j}^{\text{eff}}$,    $\sigma_{j}$    and    $L_{j}$    and
$\kappa_{j}$ is not straightforward because of the hyperbolic functions
involved. Developing the low- and high-frequency asymptotes of this equation
may help understand the roles played by the different mechanical, hydraulic and
structural properties on the seismoelectric response.

\subsection{Low-frequency asymptote}
Using first-order Taylor  expansions the following approximations can be obtained

\begin{equation}\label{cosh}
\frac{1-cosh(k_{j}L_{j})}{sinh(k_{j}L_{j})} \simeq k_{j}L_{j},
\end{equation}
\begin{equation}\label{coth}
k_{j}L_{j}\ coth(k_{j}L_{j})\simeq 1.
\end{equation}
Eqs.   (\ref{cosh})  and  (\ref{coth})  are good  approximations  when
$k_{j}L_{j} \rightarrow  0$, which occurs when  $\omega \rightarrow 0$.
Replacing  these  first-order  approximations  in Eq. (\ref{phi_cero_2}),
we get the low-frequency asymptote for the electrical potential in the center of the sample
\begin{equation}\label{phi_cero_3}
\varphi^{LF}_{z=0}(\omega)= \frac{\frac{i\omega \Delta P}{2} 
\left( \beta_1-\beta_2\right) \sum_{j=1}^{2} 
\frac{\bar{Q}_{v,j}^{\text{eff}}L_{j}} {\sigma_{j}}}
{\sum_{j=1}^{2}\frac{N_j}{L_{j}}}.
\end{equation}

\subsection{High-frequency asymptote}

In order to explore the high-frequency  behaviour of $\varphi (0)$, we use the following
approximations which are  valid  when $k_{j}L_{j}\rightarrow
\infty$
\begin{equation}\label{omega_inf_1}
\frac{1-cosh(k_{j}L_{j})}{sinh(k_{j}L_{j})}\simeq -1,
\end{equation}
\begin{equation}\label{omega_inf_2}
coth(k_{j}L_{j})\simeq 1.
\end{equation}
Replacing these expressions in (\ref{phi_cero_2}) we obtain for the high-frequency
asymptote
\begin{equation}\label{phi_inf_1}
\varphi^{HF}_{z=0}(\omega) = \frac{-i\omega \Delta P \left(\beta_1-
\beta_2\right) \sum_{j=1}^{2}\frac{\bar{Q}_{v,j}^{\text{eff}}}
{\sigma_{j}k_{j}}}
{\sum_{j=1}^{2}N_{j}k_{j}}.
\end{equation}
The expression for $k_{j}$ as  a function of $\omega$ can be
obtained by combining Eqs.(\ref{diffusivity}), (\ref{diff_length}) and
(\ref{k})
\begin{equation}\label{k_j}
k_{j}=\sqrt{\frac{i\omega \eta}{\kappa_{j}N_{j}}}.
\end{equation}
Then, replacing (\ref{k_j}) in (\ref{phi_inf_1}) one obtains
\begin{equation}\label{phi_inf_2}
\varphi^{HF}_{z=0}(\omega) = \frac{-\Delta P \left(\beta_1-
\beta_2\right) \sum_{j=1}^{2}\frac{\bar{Q}_{v,j}^{\text{eff}}}
{\sigma_{j}}\sqrt{\kappa_{j}N_{j}}}
{\sum_{j=1}^{2}\sqrt{\frac{N_{j}}{\kappa_{j}}}}.
\end{equation}

\label{lastpage}

\end{document}